\documentstyle[emulateapj,apjfonts]{article}

\newcommand{\simlt}{\mathrel{\hbox to 0pt{\lower 3.5pt\hbox{$\mathchar"218$}\hss}
      \raise 1.5pt\hbox{$\mathchar"13C$}}}
\newcommand{\simgt}{\mathrel{\hbox to 0pt{\lower 3.5pt\hbox{$\mathchar"218$}\hss}
      \raise 1.5pt\hbox{$\mathchar"13E$}}}
\newcommand{\bout}{B_{\rm out}}
\newcommand{\ceff}{C_{\rm eff}}

\begin{document}

\title{A backwards  approach to 
the formation of disk galaxies\\ I. Stellar and gas content}
\author{Ignacio Ferreras \& Joseph Silk}
\affil{Nuclear \& Astrophysics Lab. Keble Road, Oxford OX1 3RH, England, UK}
\authoremail{ferreras,silk@astro.ox.ac.uk}
\slugcomment{(Accepted for publication in  the {\it Astrophysical Journal})}

%\shorttitle{A backwards  approach to the formation of disk galaxies}
%\shortauthors{I.~Ferreras \& J.~Silk}

\begin{abstract}
A simple chemical enrichment code is described where the two
basic mechanisms driving the evolution of the ages and
metallicities of the stellar populations are the star 
formation efficiency and the fraction of gas ejected from
the galaxy. Using the observed 
Tully-Fisher relation in different passbands as a constraint,
it is found that a steep correlation between the
maximum disk rotational velocity ($v_{\rm ROT}$)
and star formation efficiency ($\ceff$) must exist 
--- $\ceff\propto v_{\rm ROT}^4$ --- either for
a linear or a quadratic Schmidt law. 
Outflows do not play a major role. This result is in contrast
with what we have found for early-type systems, where the 
Faber-Jackson constraint
in different bands allows a significant range of outflows
and requires a large star formation efficiency regardless
of galaxy mass. The extremely low efficiencies found at 
low masses translate into a large spread in the distribution
of stellar ages in these systems, as well as a large gas
mass fraction independently of the star formation law. The
model predictions are consistent with  the star formation 
rates in low-mass local galaxies. However, our predictions for
gas mass are in apparent conflict with the estimates of atomic
hydrogen content observed through the flux of the 21cm
line of \ion{H}{1}. The presence of large masses of cold 
molecular hydrogen --- especially in systems with low mass
and metallicity --- is predicted, up to ratios 
M(H$_2$)/M(HI)$\sim 4$, in agreement with a recent tentative 
detection of warm H$_2$. The redshift evolution of disk galaxies is
explored, showing that a significant change in the 
{\sl slope} of the Tully-Fisher relation ($L\propto v_{\rm ROT}^\gamma$) 
is expected because of the different age distributions 
of the stellar components in high and low-mass disk galaxies.
The slope measured in the rest frame $B, K$-bands is found to change 
from $\gamma_B \sim 3, \gamma_K \sim  4$ at $z=0$ up to $\sim 4.5, 5$ at $z\sim 1$, 
with a slight dependence on formation redshift.
\end{abstract}


%%%%%%%%%%%%%%%%%%%%%%%%%%%%%%%%%%%%%%%%%%%%%%%%%%%%%%%%%%%%%%%%%%%%%%
\section{Introduction}
The process of galaxy formation can be divided into two distinct
and equally important histories. On the one hand, the dynamical
history of galaxies is crucial in understanding the Hubble
sequence (or other morphological classifications) that we see now
as well as its evolution with redshift. On the other hand, the
star formation history allows us to trace the baryonic matter,
its conversion from gas to stars and the feedback from stellar 
evolution in the processing of subsequent generations of stars.
Disk galaxies are assumed to be the building blocks
from which the rest of the galactic ``zoo'' we see around us
has been assembled. The standard picture of disk galaxy formation
involves the infall and cooling of primordial gas onto the 
centers of dark matter halos and its later conversion into 
stars (White \& Rees 1978). A multitude of models have been 
published, describing this process in varying degrees of
detail, following either a dynamical approach, where gravity
plays the dominant role in the properties of galaxies, or a
spectrophotometric approach, where the key issue is the highly
complex and so far unsolved problem of star formation. 
The former approach usually over-simplifies the luminous
properties of galaxies by assuming a fixed value of the
mass-to-light ratio, tying the properties of the observed parts
of galaxies to the behavior of the much larger halos in which 
the luminous matter is embedded. On the other hand, 
models based on the evolution
of the stellar populations lack the dynamical information
needed in order to quantify the merger tree. Semi-analytic
models (e.g. Kauffmann et al. 1999; Baugh et al. 1998) attempt 
to fill in the gap between these two approaches at the expense
of a large and complicated set of parameters.

We follow a spectrophotometric approach, defining
a model that traces both the evolution of gas and metal content 
in a simple star formation scenario reduced to infall of
primordial gas that fuels star formation via a simple power
law dependence on gas content. The model allows for
outflows of gas driven out
of the galaxy, whose cause could either be stellar (feedback
from supernova-driven winds) or dynamical (e.g. merging or 
harassment). In this paper we study the evolution
of the gas and stellar content in disk galaxies, ignoring
size evolution. In a future paper (Ferreras \& Silk, 
in preparation) we use the same model to explore the evolution
of sizes and surface brightnesses along with the  associated effects
on the selection function of current surveys at moderate and
high redshift.

The first part of this paper (\S\S 2,3,4) describes the model
as well as the technique used in order to constrain the
volume of parameter space sampled. In \S5 we discuss the
star formation efficiencies predicted by the model. \S6 explores 
the issue of gas content in disk galaxies and  speculates on the idea
that a large fraction of gas could be found in the form of
cold H$_2$. \S7 explains the redshift evolution predicted
by our model for the star formation rates as well as for the
Tully-Fisher relation. Finally \S8 is a brief summary of
our results with a discussion of the most interesting issues, 
regarding disk galaxy formation, raised by our work.

%%%%%%%%%%%%%%%%%%%%%%%%%%%%%%%%%%%%%%%%%%%%%%%%%%%%%%%%%%%%%%%%%%%%%
\section{Modelling star formation in disk galaxies}
We follow a chemical enrichment code similar to the
one described in more detail in Ferreras \& Silk (2000a,b), 
where the formation and evolution of the stellar populations 
is parametrized by a range of star 
formation efficiencies ($\ceff$) and outflows ($\bout$). 
The model only traces the cold gas component 

%%%%%%%%%%%%%%%%%%%%%%%%%%%%%%%%%%%%%%%%%%%%%%%%%%%%%%%%%%%%%
%%%%%%%%%%%%%%%%%%%%%%%%   FIG. 1   %%%%%%%%%%%%%%%%%%%%%%%%%
%%%%%%%%%%%%%%%%%%%%%%%%%%%%%%%%%%%%%%%%%%%%%%%%%%%%%%%%%%%%%

\centerline{\null}
\vskip+3.3truein
\includegraphics{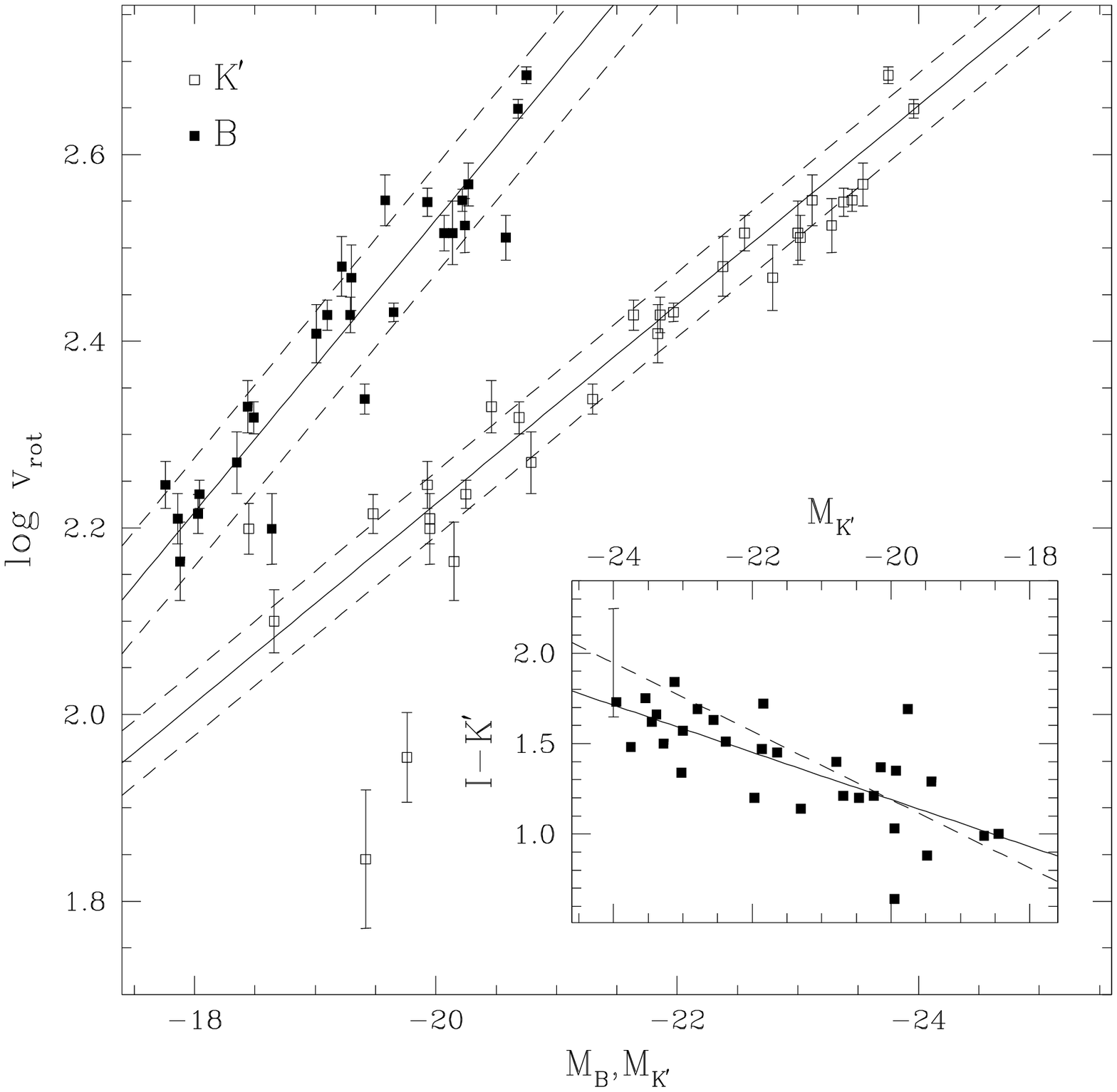}
\figcaption[fs_f1.eps]{The Tully-Fisher relation in different broadbands
	from the sample of disk galaxies in the Ursa Major cluster
	(Verheijen 1997). For clarity purposes we have plotted
	only $B$ and $K^\prime$ data, although the analysis
	presented here uses the full range of filters:
	$\{B,R,I,K^\prime\}$. The solid and dashed lines represent
	the linear fits and scatter used as constraint.
	The inset shows the $I-K^\prime$ color-magnitude	
	relation inferred by comparing the Tully-Fisher relation
	in $I$ and $K^\prime$ bands. The
	linear fit of the $I-K^\prime$ vs $M_{K^\prime}$
	color-magnitude relation from the data (solid line) 
	and from de~Grijs \& Peletier (1999; dashed line) using 
	a completely different sample of disk galaxies is
	consistent within the scatter of the data, shown as
	an error bar.\label{f1}}
\vskip+0.2truein

%%%%%%%%%%%%%%%%%%%%%%%%%%%%%%%%%%%%%%%%%%%%%%%%%%%%%%%%%%%%%

\noindent
that gives rise to star formation, with instantaneous mixing of the
ejected gas from stars and the cold gas component being
assumed. The gas mass density ($\mu_g$) and metallicity
($Z_g$) obey a simple set of chemical enrichment
equations following the formalism of Tinsley (1980):
\begin{equation}
\frac{d\mu_g}{dt}=-\psi(t)+(1-\bout)E(t)+f(t),
\end{equation}
where $\psi(t)$ is the star formation rate, $f(t)$ the gas
infall rate, $E(t)$ the gas mass ejected from stars of all
masses, and $\bout$ is the fraction of gas ejected from the
galaxy. Since this paper treats galaxies as unresolved objects,
and we do not assume a threshold in density for the star formation
rate, we can always interchange mass densities and total masses.
The evolution of the metallicity follows a similar equation:
\begin{equation}
\frac{d(Z_g\mu_g)}{dt}=-Z_g(t)\psi(t) + Z_ff(t) + (1-\bout)E_Z(t),
\end{equation}
where $Z_f$ is the metallicity of the infalling gas (assumed
to be primordial in this paper), and $E_Z(t)$ is the mass in 
metals ejected from stars. Infall of primordial gas fuels 
star formation via a Schmidt-type law (Schmidt 1959):
\begin{equation}
\psi(t) = \ceff \mu_g^n(t);
\end{equation}
where $\ceff$ is the star formation efficiency parameter.
We assume either a linear ($n=1$) or a quadratic law ($n=2$)
as extreme examples. The power law index that best fits observational
data ($n=1.4$) lies between these two values (Kennicutt 1998a).
In the instantaneous recycling approximation --- for which
the stellar lifetimes are assumed to be either zero or
infinity, depending on whether the stellar mass is 
greater or less than some mass threshold ($M_0$) --- the star
formation rate can be solved analytically (e.g. Tinsley 1980). 
Taking $n=1$:
\begin{equation}
\psi(t)=\ceff\int_0^t ds f(t-s)\exp(-s/\tau_g),
\end{equation}
\begin{equation}
\tau_g^{-1} = \ceff\big[ 1-(1-\bout)R\big];
\end{equation}
where $R$ is the returned mass fraction from stars with
masses $M>M_0$. Hence, the inverse of $\ceff$ is roughly 
the time (in Gyr) needed to process gas into stars. The 
comparison between the efficiency defined here ($\ceff$)
and observed efficiencies --- such as the parameter $A$ defined in
Kennicutt (1998b) with respect to surface densities rather 
than total masses, i.e. $A=\Sigma_\psi /\Sigma^n_{\rm gas}$ 
--- is straightforward without a star formation threshold, but 
gets rather complicated with such a threshold, which causes 
$\ceff$ to lie systematically below the observed efficiencies 
by a factor given by the ratio between the surface of the disk 
with density above threshold with respect to the total area. 
If we assume a universal star formation law, one would expect 
a systematic offset between observed star formation efficiencies 
and model one-zone efficiencies such as our $\ceff$.

The infall rate that fuels star formation is assumed to have
a Gaussian profile peaked at an epoch described by a formation
redshift parameter ($z_F$), with two different timescales
for ages earlier or later than the time corresponding to that
formation redshift, namely:
\begin{equation}
f(t)\propto \left\{ 
\begin{array}{lr}
\exp [-(t-t(z_F))^2/2\tau_1^2] & t < t(z_F)\\
\exp [-(t-t(z_F))^2/2\tau_2^2] & t > t(z_F).\\
\end{array}
\right.
\end{equation}

For $t<t(z_F)$ we assume a short infall timescale --- $\tau_1=0.5$~Gyr
--- which results in a prompt enrichment of the interstellar
medium and thereby circumvents the G~dwarf problem. Any model of chemical
enrichment must avoid the over-production of low metallicity
stars by assuming either a non-monotonically decreasing
star formation rate (as in this paper) or initially non-primordial
gas (pre-enrichment). 
For $t>t(z_F)$ the infall timescale ($\tau_2$)
is a rough label for disk type: early-type disks will be
associated with short timescales ($\tau_2\sim 1-2$ Gyr), 
whereas late-type disks, with more extended star formation 
histories, correspond to larger values of $\tau_2$. 
The Initial Mass Function (IMF) used is a hybrid between 
a Salpeter (1955) and a Scalo (1986) IMF. 
The high-mass end follows 
a Salpeter power law, with upper-mass cutoff at 
$60 M_\odot$, whereas the low-mass end --- truncated 
at $0.1 M_\odot$ --- follows the Scalo IMF.
Stellar lifetimes follow a broken power law fit to the
data from Tinsley (1980) and Schaller et al. (1992).
We refer the reader to Ferreras \& Silk (2000a,b) for more 
details about the chemical enrichment code. 

The outflow parameter ($\bout$) is used to quantify the amount
of gas ejected out of stars and that leaves the galaxy
and does not contribute to the processing of the next generation
of stars. This mechanism plays an important role in the
mass-metallicity correlation as originally suggested by Larson
(1974): massive galaxies are expected to retain most of the 
gas, thereby increasing the average metallicity, whereas
the shallow gravitational wells of low-mass galaxies cannot
prevent enriched gas from being ejected from the galaxy,
resulting in lower average metallicities. It is a well-known
fact that the color range found in early-type systems 
is mostly driven by metallicity 

%%%%%%%%%%%%%%%%%%%%%%%%%%%%%%%%%%%%%%%%%%%%%%%%%%%%%%%%%%%%%
%%%%%%%%%%%%%%%%%%%%%%%%   FIG. 2   %%%%%%%%%%%%%%%%%%%%%%%%%
%%%%%%%%%%%%%%%%%%%%%%%%%%%%%%%%%%%%%%%%%%%%%%%%%%%%%%%%%%%%%

\centerline{\null}
\vskip+3.3truein
\includegraphics{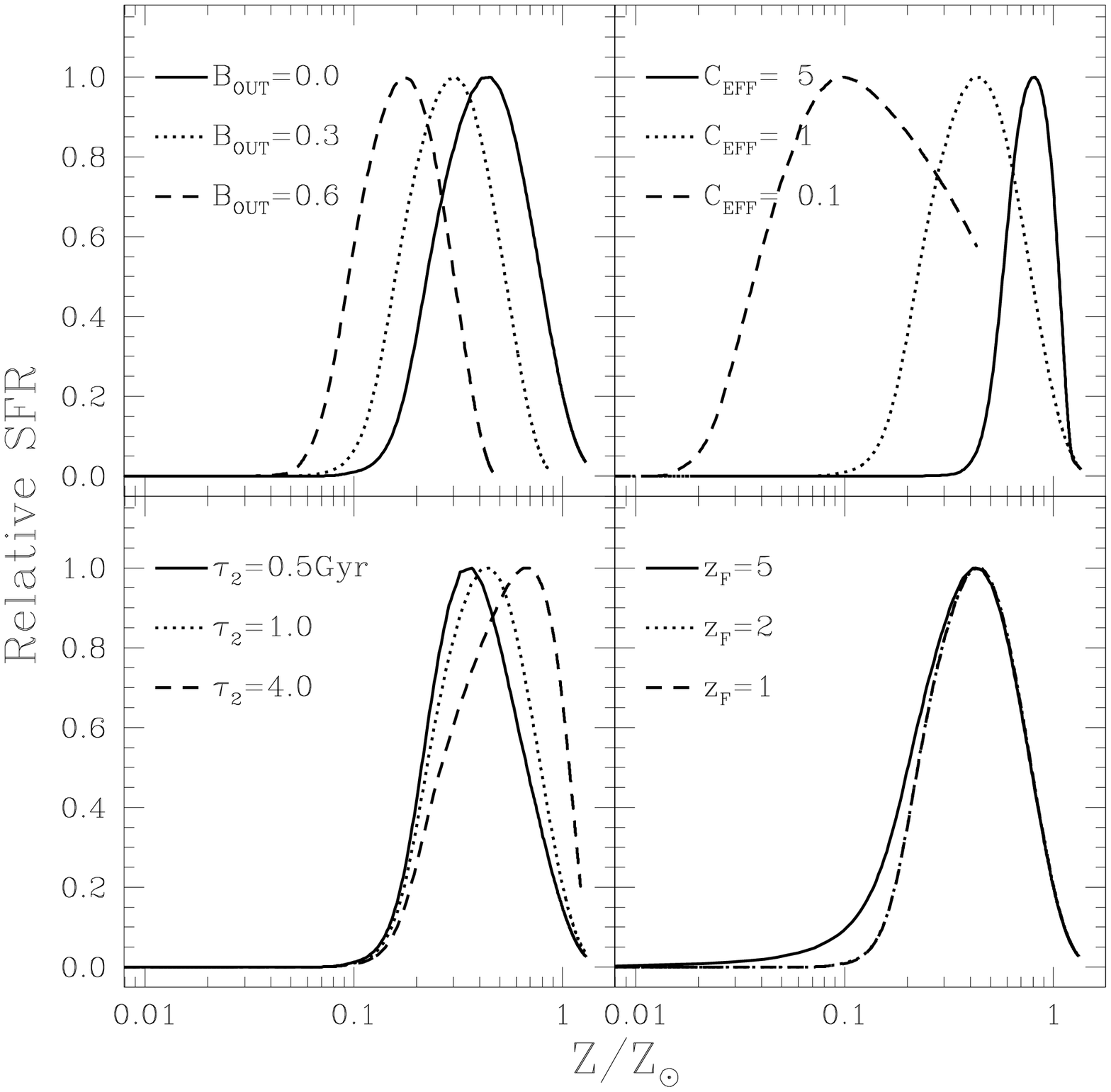}
\figcaption[fs_f2.eps]{Histogram of metallicities of the 
	stellar populations following the chemical enrichment 
	code described here (using a linear Schmidt-type law) 
	for a range of outflows ($\bout$; {\sl top left});
	star formation efficiencies ($\ceff$; {\sl top right});
	infall timescale ($\tau_2$, {\sl bottom left}); and
	epoch at peak of infall, given as a formation
	redshift ($z_F$). The fiducial set of parameters 
	is ($\bout$,$\ceff$,$\tau_2$,$z_F$)=(0,1,1~Gyr,2).\label{f2}}
\vskip+0.2truein

%%%%%%%%%%%%%%%%%%%%%%%%%%%%%%%%%%%%%%%%%%%%%%%%%%%%%%%%%%%%%

\noindent
(e.g. Kodama et al. 1998), 
and can be explained by a correlation between the outflow 
parameter ($\bout$) and the total mass of the system 
(Ferreras \& Silk 2000b).

The chemical enrichment process in galaxies is hence
described in this model with a set of five parameters:
$(\tau_1,\tau_2,z_F,\ceff,\bout)$. Only the early-infall
timescale --- $\tau_1$ --- is fixed to 0.5 Gyr throughout, 
whereas the other four parameters
are allowed to vary over a large range of values. 
Every set of parameters chosen describes a star formation
history ($\psi(t)$, $Z_g(t)$) which can be used to convolve
simple stellar populations in order to find the composite
spectral energy distribution of a galaxy whose formation 
process corresponds to that choice of parameters. We
use the latest population synthesis models from Bruzual
\& Charlot (in preparation), which span the range of
metallicities $\frac{1}{50} \leq Z/Z_\odot \leq \frac{5}{2}$
and include all phases of stellar evolution, from the
zero-age main sequence to supernova explosions for progenitors
more massive than $8 M_\odot$, or to the end of the white dwarf
cooling sequence for less massive progenitors. The uncertainties
present in population synthesis models (Charlot, Worthey 
\& Bressan 1996) complicate the determination of absolute values
for parameters such as stellar ages or star formation efficiencies.
This paper aims at showing {\sl relative} values of efficiencies
or outflow parameters between galaxies of different masses.

The predicted spectral energy distribution obtained for
each point scanned in this large volume of parameter
space is matched against observed data so that a map
of the parameter range compatible with observations
can be drawn. Here we use the Tully-Fisher relation of disk 
galaxies in different passbands as a constraint, shown in 
Figure~1. The data involve multiband $\{B, R, I, K^\prime\}$ 
photometry of a sample of disk galaxies in the Ursa Major 
cluster (Verheijen 1997). Section~4 describes the comparison 
in more detail.

%%%%%%%%%%%%%%%%%%%%%%%%%%%%%%%%%%%%%%%%%%%%%%%%%%%%%%%%%%%%%
%%%%%%%%%%%%%%%%%%%%%%%%   FIG. 3   %%%%%%%%%%%%%%%%%%%%%%%%%
%%%%%%%%%%%%%%%%%%%%%%%%%%%%%%%%%%%%%%%%%%%%%%%%%%%%%%%%%%%%%

\centerline{\null}
\vskip+3.3truein
\includegraphics{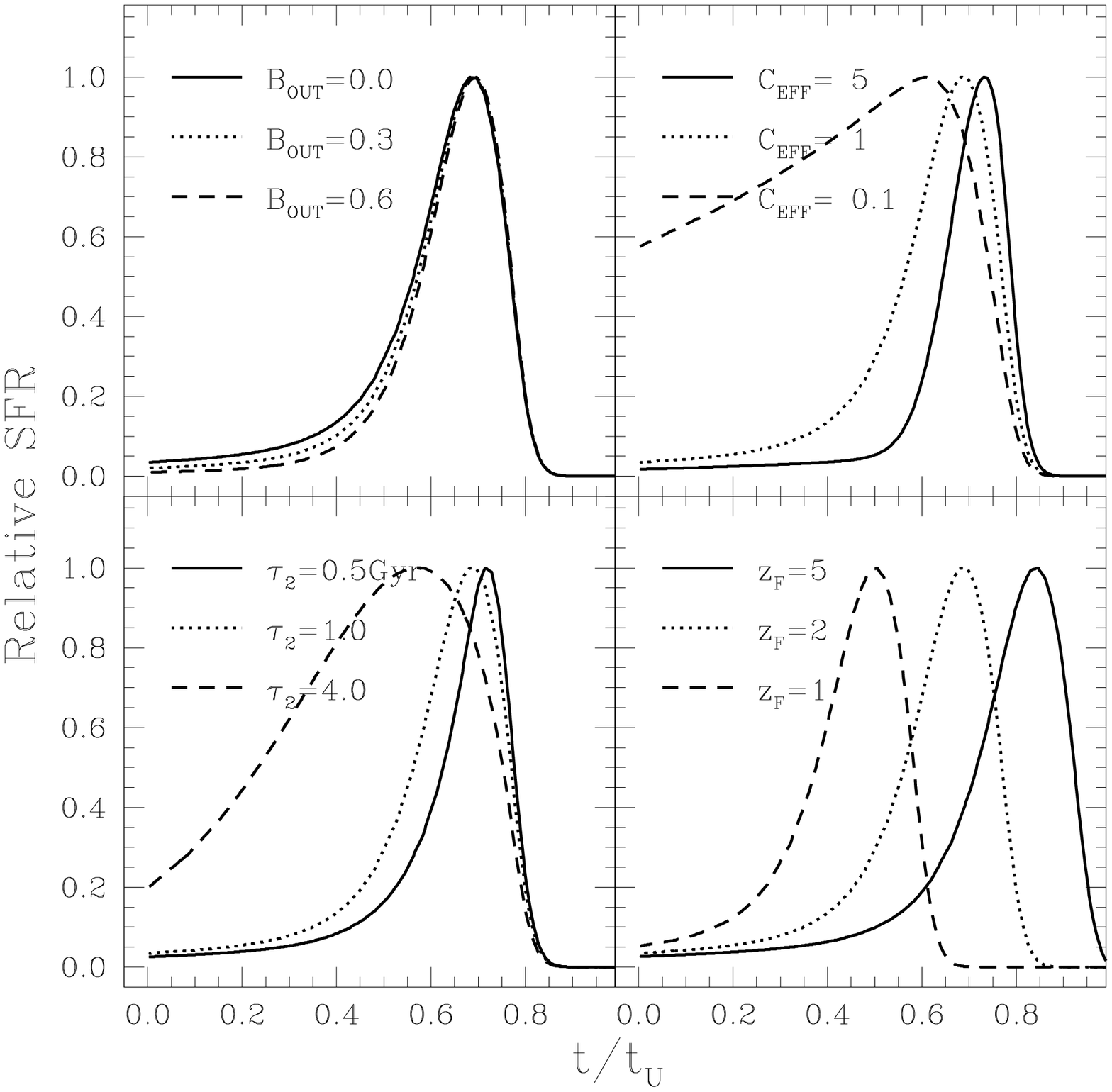}
\figcaption[fs_f3.eps]{Age histogram of the stellar populations for the
	same range of parameters as in Figure~2. 
	The ages are given with respect to the age of the universe 
	at zero redshift: $t_U$, which is 13.5~Gyr for a
	$H_0=70$ km s$^{-1}$Mpc$^{-1}$; $\Omega_m=0.3$, 
	$\Lambda$-dominated flat cosmology, used throughout 
	the paper.\label{f3}}
\vskip+0.2truein

%%%%%%%%%%%%%%%%%%%%%%%%%%%%%%%%%%%%%%%%%%%%%%%%%%%%%%%%%%%%%

%%%%%%%%%%%%%%%%%%%%%%%%%%%%%%%%%%%%%%%%%%%%%%%%%%%%%%%%%%%%%%%%%%%%%
\section{Outflows versus Star Formation Efficiencies}

In order to visualize the role of every parameter
used in the chemical enrichment model presented here, 
we show in Figures~2 and 3 the metallicity and age 
distribution, respectively, for a range of outflows
($\bout$); star formation efficiencies ($\ceff$) 
and infall parameters ($\tau_2$, $z_F$). The fiducial 
set chosen is ($\bout$,$\ceff$,$\tau_2$,$z_F$)=
(0,1,1~Gyr,2). In each panel, one of the parameters is
allowed to vary, while keeping the others fixed. Notice the
effect of the early-infall timescale ($\tau_1$) 
on the metallicity distribution, which avoids the 
over-production of low metallicity stars. If we had set the
timescale $\tau_1$ to zero, the metallicity histogram 
would have been a monotonically decreasing function, 
peaked at zero metallicity. Closed-box models with constant 
star formation efficiencies and a standard initial mass function
always yield monotonically
decreasing star formation rates that over-produce low metallicity
stars which contradict observations, both in our local 
stellar census (Rocha-Pinto \& Maciel 1996) as well as in the 
integrated photometry of bulge-dominated galaxies 
(Worthey, Dorman \& Jones 1996). Models with pre-enriched gas
fuelling star formation generate similar distributions
although shifted towards higher abundances. The photometric
predictions of pre-enrichment models can be made compatible with
the observations. However, the metallicity distribution
--- a monotonically decreasing function with $Z$ with 
the maximum at the pre-enriched metallicity --- is in disagreement 
with the observed metallicities of stars in the Milky Way
(e.g. Rocha-Pinto \& Maciel 1996). However, see Rich (1990) and
Ibata \& Gilmore (1995), who find an agreement between the
prediction of a closed-box model and the distribution in 
the galactic bulge.

The top left panels of Figures~2 and 3 show that a range 
in outflows --- with all the other parameters fixed --- 
does not generate significantly different age 
distributions. The net effect of a change in the amount 
of gas ejected from the galaxy is a shift 

%%%%%%%%%%%%%%%%%%%%%%%%%%%%%%%%%%%%%%%%%%%%%%%%%%%%%%%%%%%%%
%%%%%%%%%%%%%%%%%%%%%%%%   FIG. 4   %%%%%%%%%%%%%%%%%%%%%%%%%
%%%%%%%%%%%%%%%%%%%%%%%%%%%%%%%%%%%%%%%%%%%%%%%%%%%%%%%%%%%%%

\centerline{\null}
\vskip+3.3truein
\includegraphics{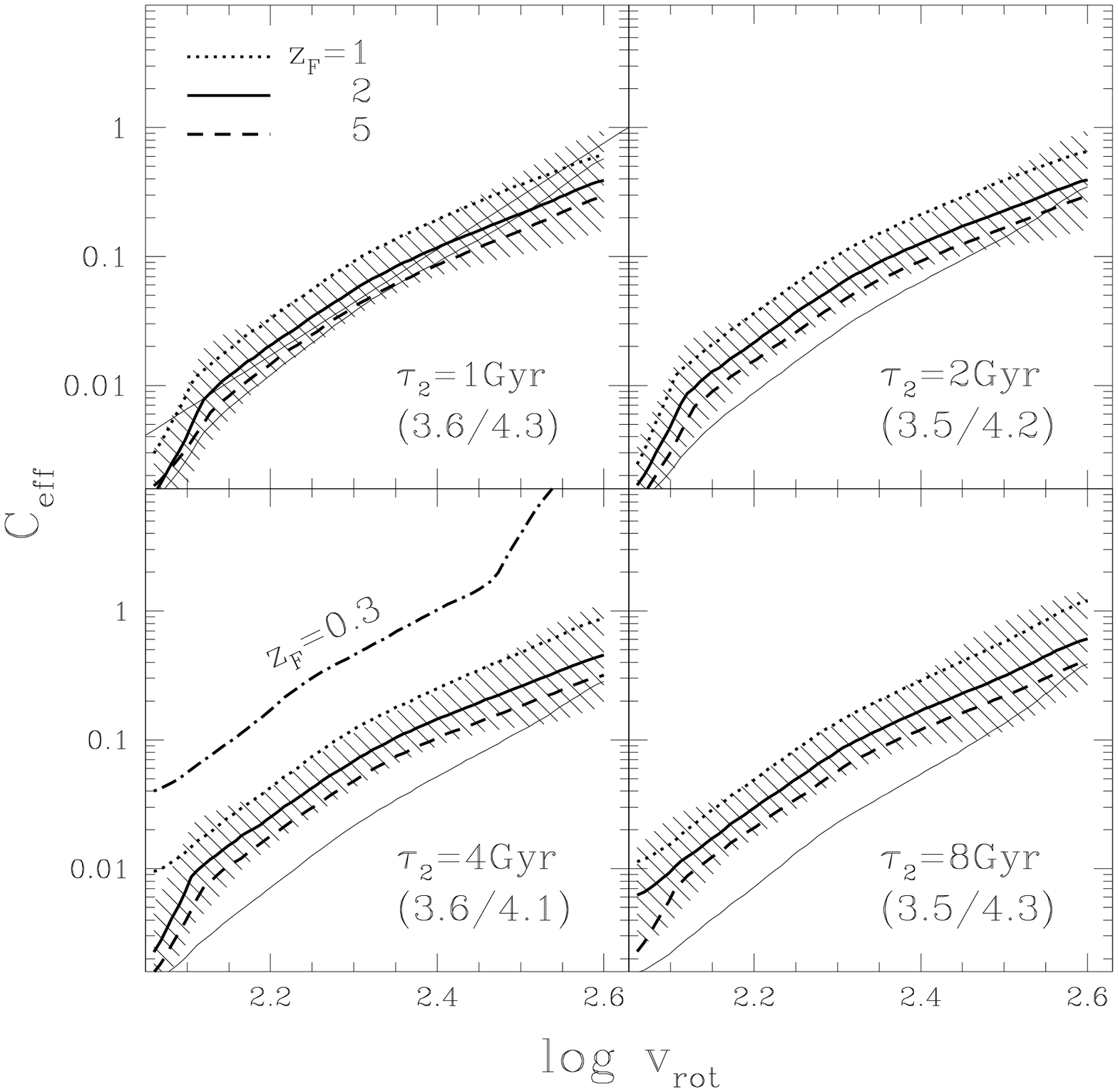}
\figcaption[fs_f4.eps]{Range of star formation efficiencies ($\ceff$) as 
	a function of rotational velocity obtained using
	the multi-band Tully-Fisher relation as constraint
	with a linear Schmidt-type law.
	The thick dotted, solid and dashed lines correspond to 
	formation redshifts of $z_F=1,2$ and $5$,
	respectively. The shaded area represents the 90\%
	confidence level in the $\chi^2$ fit for $z_F=2$. The thin
	solid line gives the range of efficiencies in the model
	with $z_F=2$ using a quadratic Schmidt-type law.
	The numbers in parentheses for every infall timescale
	give the slope $\Delta\log\ceff /\Delta\log v_{\rm ROT}$
	for $z_F=2$, using a linear and quadratic Schmidt law, 
	respectively. The dot-dash line in the bottom-left
	panel shows how a very late formation redshift ($z_F=0.3$) 
	results in an increase of the star formation 
	efficiency.\label{f4}}
\vskip+0.2truein

%%%%%%%%%%%%%%%%%%%%%%%%%%%%%%%%%%%%%%%%%%%%%%%%%%%%%%%%%%%%%

\noindent
in the average 
metallicity. Hence, the assumption that an outflow range is 
the only mechanism driving the luminosity sequence of 
galaxies results in a pure metallicity sequence. On the other
hand, a range of star formation efficiencies generates
distributions with different ages and metallicities. The 
efficiency parameter sets the clock rate at which stars 
are being formed from the infalling gas (equations 4 and 5). 
Hence, systems with lower efficiencies generate a larger 
age spread, and lower
average metallicities. Finally, the parameters controlling infall
($\tau_2$ and $z_F$) simply shift the age distributions without
a significant change in the metallicities, as expected,
since $\bout$ and $\ceff$ are the only parameters
driving the chemical enrichment process: it is possible to 
re-define the time variable as $s\equiv (t-t(z_F))/\tau_2$ and 
solve similar chemical enrichment equations to (1) and (2) 
with respect to variable $s$, without any significant
dependence on the infall parameters ($\tau_2$ and $z_F$).

A comparison of the age and metallicity histograms with the 
observed mass-metallicity correlation found in disk galaxies 
(Zaritsky, Kennicutt \& Huchra 1994) suggests that either 
outflows and/or star formation efficiencies must be strongly
correlated with galaxy mass. However, one can not determine
{\sl a priori} whether the observed color range found in disks
is purely driven by outflows (i.e. a metallicity sequence) 
or by efficiency (i.e. a mixed age $+$ metallicity sequence).

%%%%%%%%%%%%%%%%%%%%%%%%%%%%%%%%%%%%%%%%%%%%%%%%%%%%%%%%%%%%%%%%%%%%%
\section{The Tully-Fisher relation as a constraint}

The work described here follows a ``backwards'' approach
in that we apply correlations between observables in 
local galaxies as a constraint and use the chemical 
enrichment code in order 

%%%%%%%%%%%%%%%%%%%%%%%%%%%%%%%%%%%%%%%%%%%%%%%%%%%%%%%%%%%%%
%%%%%%%%%%%%%%%%%%%%%%%%   FIG. 5   %%%%%%%%%%%%%%%%%%%%%%%%%
%%%%%%%%%%%%%%%%%%%%%%%%%%%%%%%%%%%%%%%%%%%%%%%%%%%%%%%%%%%%%

\centerline{\null}
\vskip+3.3truein
\includegraphics{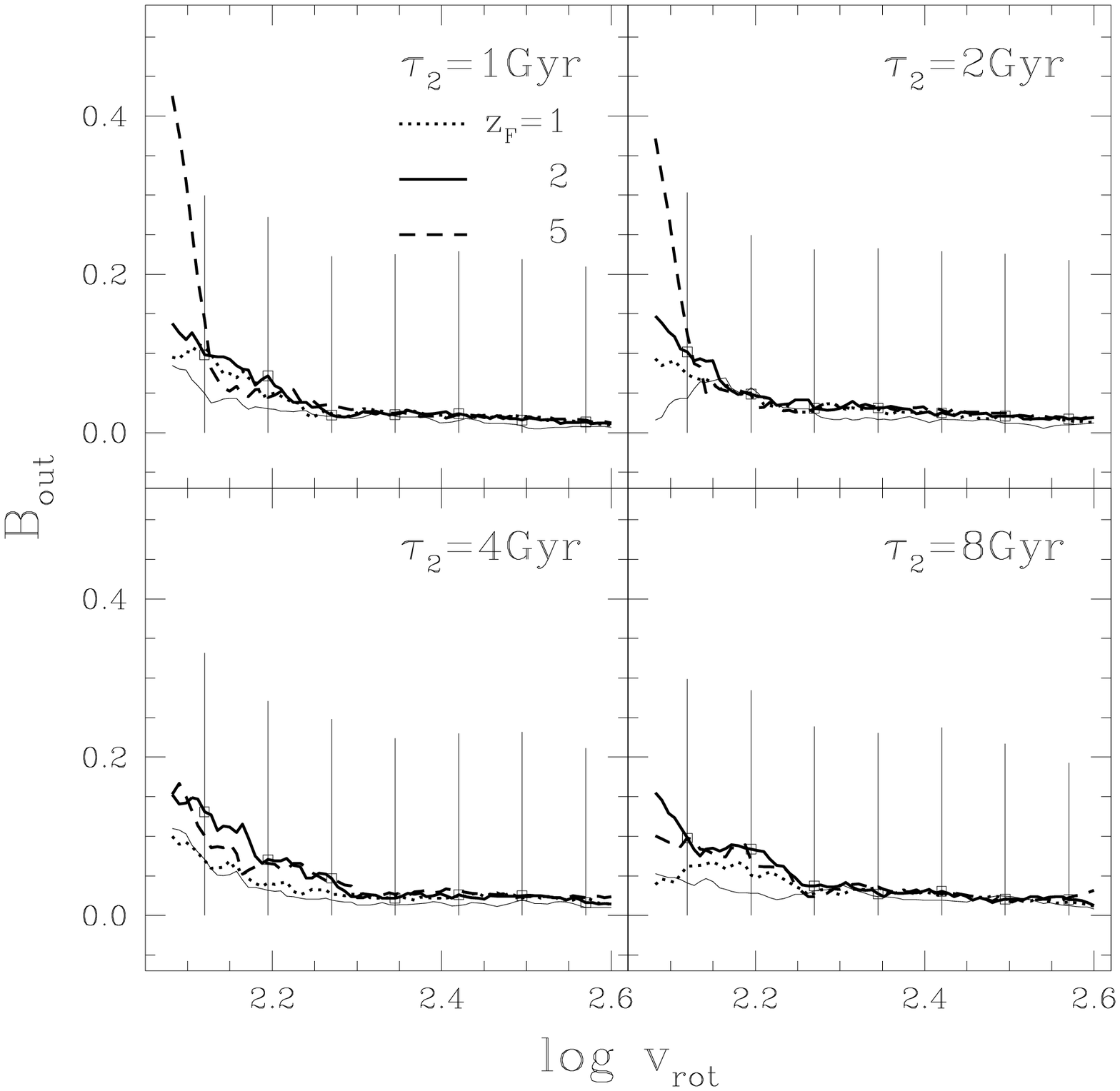}
\figcaption[fs_f5.eps]{Range of outflows as a function of rotational
	velocity for four different infall timescales
	(linear Schmidt law).
	The error bars are the 90\% confidence levels
	in the $\chi^2$ analysis for the $z_F=2$ case.
	The dotted, solid and dashed lines represent
	different formation redshifts ($z_F=1,2$ and $5$,
	respectively). The thin solid line is the
	result for a quadratic Schmidt-type law with $z_F=2$.\label{f5}}
\vskip+0.2truein

%%%%%%%%%%%%%%%%%%%%%%%%%%%%%%%%%%%%%%%%%%%%%%%%%%%%%%%%%%%%%

\noindent
to evolve the system backwards 
in time. For disk galaxies the obvious constraint to 
be used is the Tully-Fisher relation (Tully \& Fisher 1977):
this shows a tight correlation between the disk rotation velocity 
--- which is a measure of the ratio between the mass and 
the size of the galaxy --- and the absolute luminosity. 
The scatter of this correlation is rather small 
($\sim 0.3$ mag) which allows it to be used as a distance 
indicator with uncertainties as low as 10\%, notwithstanding
unknown systematic effects  (e.g. Jacoby et al. 1992).
The small scatter hides a fundamental connection between
the dynamical evolution of disk galaxies and their
star formation histories. The analogous correlation for
bulges or early-type systems --- the Faber-Jackson relation
(Faber \& Jackson 1976) between luminosity and velocity 
dispersion of the spheroid --- has a larger scatter, 
a clear sign of a more complex dynamical evolution. 

The conventional scenario for disk formation via 
infall of gas onto the centers of dark matter 
halos gives a qualitative
idea of the trend followed by the Tully-Fisher relation:
massive disks (which have faster rotation velocities) will 
create more stars, thereby having higher luminosities than
low-mass systems. However, a more detailed analysis of the
Tully-Fisher relation --- including disk sizes which 
of course affect the measured rotation velocity --- 
shows that the interpretation of the
slope and zero point of the Tully-Fisher relation is 
more complicated. Many dynamical
models assume constant mass-to-light ratios ($M/L$), 
an incorrect assumption unless only stellar populations 
of the same ages dominate the light from all disks.
The Tully-Fisher relation observed through near-infrared
passbands minimizes the effect of the chemical enrichment
history, and so $K$-band is the preferred filter in
order to use the correlation as a distance indicator.
Rather than exploring a suitable way of finding a 
``distilled'' Tully-Fisher relation that is independent
of the formation process, we will use the correlation 
in different filters in order to constrain the star 
formation history of disk galaxies. There are many 
samples of disk galaxies for which rotation curves
have been observed using different techniques such as
optical emission lines (Mathewson \& Ford 1996)
or the 21~cm line of atomic hydrogen 
(Giovanelli et al. 1997). We use the sample of disk galaxies 
in the Ursa Major cluster (Tully et al. 1996; Verheijen 1997)
imaged in several filters: $B$, $R$, $I$ and $K^\prime$.
The rotation curves were determined with the 21~cm line of H~I.

For a choice of parameters describing the star formation
history: ($\bout$, $\ceff$, $\tau_2$, $z_F$), we solve 
equations (1) and (2) describing the evolution of the
gas mass and metallicity and use this to convolve
simple stellar populations from the latest models
of Bruzual \& Charlot (in preparation). 
We explore four different formation redshifts
($z_F=\{1, 2, 3, 5\}$) and four different infall 
timescales ($\tau_2 = \{1, 2, 4, 8\}$~Gyr), corresponding
to a range of galaxy types: from early-type disks
such as S0, Sa --- for $\tau_2\sim 1-2$~Gyr --- 
to late-type systems such as Sd for longer timescales.
For every pair of
values describing infall ($\{\tau_2, z_F\}$) we scan
a grid of $128\times 128$ star formation efficiencies
and outflow fractions over the following range:
\begin{center}
$-3 \leq \log\ceff \leq +1$\\
$0 \leq \bout \leq 0.7$.
\end{center}
\noindent
For every choice of parameters, the $I-K^\prime$ 
color of the un-normalized composite spectral energy 
distribution is used in order to find the absolute 
luminosity using the color-magnitude relation of the
sample of Verheijen (1997):
\begin{equation}
M_K^\prime = -\frac{(I-K^\prime) + 1.417}{0.1304},
\end{equation}
which is shown as a solid line in the inset of Figure~1.
The color-magnitude relation found by de~Grijs \& Peletier
(1999) from a subsample of the Surface Photometry Catalogue
of ESO-Uppsala galaxies is shown as a dashed 
line, in remarkable agreement with the fit from the sample
of UMa galaxies. In fact, the universality 
of the color-magnitude relation in disk galaxies is 
explored by de~Grijs \& Peletier (1999) as a plausible 
distance indicator, with accuracies deemed to be 
better than 25 \%.

%%%%%%%%%%%%%%%%%%%%%%%%%%%%%%%%%%%%%%%%%%%%%%%%%%%%
%%%      Table 1. Tully-Fisher parameters        %%%
%%%%%%%%%%%%%%%%%%%%%%%%%%%%%%%%%%%%%%%%%%%%%%%%%%%%

 \begin{table*}
   \begin{center}
   \caption{Fits to the Tully-Fisher relation: 
	$M_X^{\rm TF}=\alpha_X\log v_{\rm ROT}+\beta_X$, and scatter
	$\sigma_X$}
   \label{table-1}
   \vskip+0.1truein
   \begin{tabular}{l|ccc}\hline\hline
  Band ($X$) & $\alpha_X$ & $\beta_X$ & $\sigma_X$\\
  \hline
  $B$        & $-6.387$ & $-3.842$ & $0.370$\\
  $R$        & $-7.356$ & $-2.284$ & $0.348$\\
  $I$        & $-8.020$ & $-1.061$ & $0.394$\\
  $K^\prime$ & $-9.368$ & $+0.849$ & $0.323$\\
  \hline\hline
   \end{tabular}
   \end{center}
 \end{table*}

Once the spectral energy distribution is normalized, we
can compare the predicted Tully-Fisher relations in 
different wavebands with the observations. Table~1 shows
the result of fitting the photometry of disk galaxies
in the Ursa Major cluster (Verheijen 1997) to a 
Tully-Fisher relation in different filters 
($X=\{B, R, I, K^\prime\}$):
\begin{equation}
M_X^{\rm TF} = \alpha_X\log v_{\rm ROT} + \beta_X.
\end{equation}
We perform a two-stage linear fit with a first estimate 
of the slope and zero point by applying absolute deviations.
This first estimate is used to remove points which lie
more than one standard deviation off the linear fit. 
A least squares fit is 
subsequently applied to the remaining points. This technique 
prevents outliers from contributing significantly to the fit.
These fits allow us to estimate how well different star formation
histories --- described by the parameters --- can account
for the observed Tully-Fisher relations in different
passbands. We define a $\chi^2$ by:
\begin{equation}
\chi^2\equiv\Sigma_{X=\{B, R, I, K^\prime\}}
\frac{(M_X-M_X^{\rm TF})^2}{\sigma_X^2}.
\end{equation}

%%%%%%%%%%%%%%%%%%%%%%%%%%%%%%%%%%%%%%%%%%%%%%%%%%%%%%%%%%%%%
%%%%%%%%%%%%%%%%%%%%%%%%   FIG. 6   %%%%%%%%%%%%%%%%%%%%%%%%%
%%%%%%%%%%%%%%%%%%%%%%%%%%%%%%%%%%%%%%%%%%%%%%%%%%%%%%%%%%%%%

\centerline{\null}
\vskip+3.3truein
\includegraphics{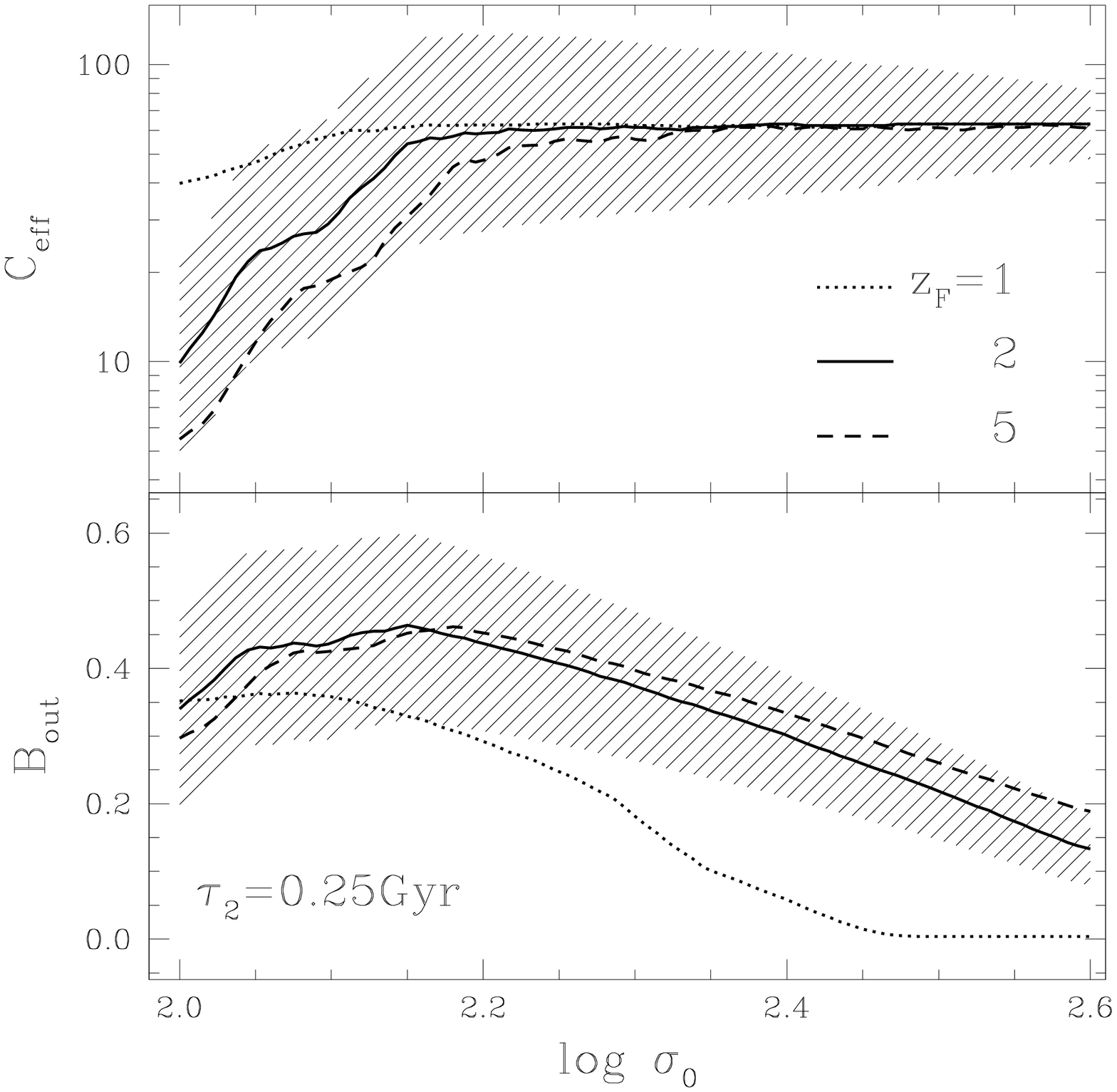}
\figcaption[fs_f6.eps]{Range of star formation efficiencies ({\sl top})
	and gas outflows ({\sl bottom}) as a function of
	central velocity dispersion ($\sigma_0$) that give the best
	fit to the chemical enrichment model constrained
	by the Faber-Jackson relation observed in Coma
	elliptical galaxies in $\{U, V, J, {\rm and} K\}$ bands. 
	The shaded area represents the 90\% confidence
	levels of the $\chi^2$ analysis performed for
	the $z_F=2$ model.\label{f6}}
\vskip+0.2truein

%%%%%%%%%%%%%%%%%%%%%%%%%%%%%%%%%%%%%%%%%%%%%%%%%%%%%%%%%%%%%

Figures~4 and 5 show the range of star formation efficiencies
and outflow fractions, respectively, allowed by the observed 
photometry. This range is shown in four panels corresponding 
to four different infall timescales (i.e. four different galaxy
types). Each panel gives the result for three different formation
redshifts (thick lines) using a linear Schmidt law ($n=1$
in equation 3). The result for a quadratic law ($n=2$) is
shown for $z_F=2$ as a thin solid line. The 90\% confidence 
level of the $\chi^2$ fit for $z_F=2$ and $n=1$ is shown as 
a shaded area in Figure~4 and as error bars in Figure~5.

%%%%%%%%%%%%%%%%%%%%%%%%%%%%%%%%%%%%%%%%%%%%%%%%%%%%%%%%%%%%%%%%%%%%%%
\section{How efficient is Star Formation~?}
Figures~4 and 5 show that a very large 
range of star formation efficiencies is needed in order
to explain the observations, whereas outflows seem to be 
unimportant in disk galaxies. Over a velocity range of
$\Delta\log v_{\rm ROT}=0.6$ dex we find a range of
star formation efficiencies corresponding to gas-to-star 
processing timescales that span over 2 decades 
between very small
disks ($v_{\rm ROT}\sim 100$ km s$^{-1}$) and massive
ones ($v_{\rm ROT}\sim 400$ km s$^{-1}$) for various
infall parameters. This result is
very robust given the assumptions and the model presented 
here, as well as the 90\% confidence level of our 
$\chi^2$ fit. While outflows could play a
role, these cannot be very important. The error 
bars shown in Figure~5 show that outflows in low-mass 
systems cannot be larger than 20\%. Furthermore, a constant
value $\bout\sim 0$, regardless of $v_{\rm ROT}$, is 
also consistent with the data. This is also in agreement
with the negligible effect of supernova-driven winds in
disks as found in hydrodynamic simulations 
(MacLow \& Ferrara 1999). 
The correlation between $\ceff$ and $v_{\rm ROT}$ does not change
significantly with respect to morphological type (roughly described
by our infall timescale $\tau_2$). This is in agreement with 
Young et al. (1996) who find no variation in the global ratio
$L(H\alpha)/M(H_2)$ for spiral galaxies with types Sa-Sc.

The range of efficiencies is consistent with the observations 
of Kennicutt (1998b),
who defines an efficiency with respect to surface
densities: $A\equiv \Sigma_\psi / \Sigma^n_{\rm gas}$.
The data fit a power law index $n=1.4\pm 0.15$,
with the efficiency ranging over a factor of $100$.
However, the observed efficiencies are systematically one order
of magnitude higher than our $\ceff$, with the lowest value
around $A\sim 0.1$ (thereby the gas consumption timescales
$A^{-1}\simlt 10$~Gyr are significantly shorter than $\ceff^{-1}$).
There are various reasons for such a discrepancy between $A$ 
and $\ceff$. We believe the most important one
to be a systematic underestimate of the amount of
molecular hydrogen, which we discuss in more detail in 
the next sections. Furthermore, the star formation rates
obtained from $H\alpha$ measurements are corrected for extinction
by a factor of 2.8 using a comparison between $H\alpha$ fluxes and
radio fluxes due to free-free emission at 3 and 6~cm 
(Kennicutt 1983). However, the radio emission is assumed to be purely
thermal. A non-thermal contribution in the radio fluxes would
imply a systematic overestimate of the star formation rates, with a
corresponding increase of $A$ with respect to $\ceff$.
Additionally, a universal star formation 
threshold in the surface gas density would result in lower
integrated efficiencies --- such as the parameter $\ceff$ ---
compared to surface-based measurements such as $A$, with the 
correction factor roughly amounting to the ratio of
the disk area over which the surface mass density is above 
threshold with respect to the total area of the galaxy.
A linear Schmidt law and a surface density threshold for 
the star formation ($\Sigma_{th}$) would imply:
\begin{equation}
\psi(t)=\ceff M_g(t)=A\int d^2r\Sigma_g(r,t)\Theta\Big( 
\Sigma_g(r,t)-\Sigma_{th}\Big) < AM_g(t),
\end{equation}
where $\Theta(x)$ is the step function. For an exponential
density profile with scale-length $h$, i.e. $\Sigma(r)=\Sigma_0\exp (-r/h)$
we find:
\begin{equation}
\ceff = A\Big( 1-e^{-x_0}\Big) < A,
\end{equation}
such that $\Sigma(r=x_0h)=\Sigma_{th}$. Hence, a threshold will
cause a systematic offset towards lower values for the integrated
star formation efficiencies.
A delay in the galaxy formation process can also yield
higher star formation efficiencies. The dot-dash line in the 
bottom-left panel of Figure~4 gives the star formation efficiencies
predicted for $z_F=0.3$, which are about one order of magnitude
higher than those for models with $z_F\geq 1$. However, this implies
the bulk of the stellar populations should have ages~$\simlt 2.5$~Gyr,
which leads to a drastic change of the luminosity function
of disk galaxies at moderate redshift that is not seen 
(e.g. Lilly et al. 1995).

The large range of star formation efficiencies shown in 
Figure~4 is in remarkable contrast with the region of parameter 
space allowed by the color-magnitude
constraint in early-type systems. The photometry of spheroids
is compatible with either a large range of star formation
efficiencies {\sl and/or} a wide range of gas ejected in 
outflows (Ferreras \& Silk 2000b). In fact, the lack of evolution 
of the slope of the color-magnitude relation of early-type 
galaxies with redshift suggests the mass sequence of early-type
galaxies is a pure metallicity sequence, where the stellar
populations have roughly similar ages regardless of galaxy mass
so that the color range observed is explained by a 
mass-metallicity correlation. Such a correlation, without a
similar mass-age connection, can only be explained by assuming
outflows are very important in low-mass systems. 
In order to test our model, we applied the same chemical
enrichment code and fitting technique to the Faber-Jackson
relation in early-type systems. We used the $U$, $V$, $J$ and $K$
photometry of elliptical galaxies in the Coma cluster
from Bower, Lucey \& Ellis (1992). The best fits are shown in 
Figure~6  for the star formation efficiency ({\sl top}) and
outflow fraction ({\sl bottom}), taking a linear Schmidt
law and three different formation redshifts: 
$z_F=\{1, 2, 5\}$.  We chose
a short infall timescale: $\tau_2=0.25$~Gyr, although
larger values were also used, leading to similar results for
$\tau_2=0.5$ and $1$ Gyr. Longer infall timescales were 
incompatible with the photometric data. The shaded area 
represents the 90\% confidence level of the $\chi^2$
fit in the $z_F=2$ case. In contrast to disk galaxies
(Figures~4 and 5), early-type systems require very high
star formation efficiencies, and are compatible with a 
significant correlation between galaxy mass and
fraction of gas ejected in outflows. This is in agreement
with the enhanced star formation efficiencies observed in 
mergers (Young et al. 1996), which are believed to be 
associated with the process of spheroid formation. A 
plausible scenario to reconcile these two very different 
trends is that
disk galaxies are dynamically ``ordered'' systems, where
gas ejection can only take place through feedback from 
supernovae, whereas early-type galaxies (or disk
bulges) have a more complicated dynamical history, which can
allow for a significant amount of gas (and metals)
being ejected from the galaxy, as expected in dwarf galaxy
mergers (Gnedin 1998). Furthermore, this
process should be correlated with the mass of the galaxy
(i.e. more gas being ejected in events involving
less massive mergers).

%%%%%%%%%%%%%%%%%%%%%%%%%%%%%%%%%%%%%%%%%%%%%%%%%%%%%%%%%%%%%%%%%%%%%
\section{Of gas and stars}

The large range of star formation efficiencies required 
by our model --- shown in Figure~4 --- implies that 
there is a large difference in gas content with respect
to galaxy luminosity. As a rough approximation, the inverse
of the star formation efficiency gives the timescale
for the processing of gas into stars (in Gyr). We can see 
in the figure
that in the interval $2.1<\log v_{\rm ROT}<2.6$
--- which corresponds to a range of $4.5$ magnitudes
in absolute $K$-band luminosity --- $\ceff$ spans 
two orders of magnitude. This means it takes
$100$ times longer to 

%%%%%%%%%%%%%%%%%%%%%%%%%%%%%%%%%%%%%%%%%%%%%%%%%%%%%%%%%%%%%
%%%%%%%%%%%%%%%%%%%%%%%%   FIG. 7   %%%%%%%%%%%%%%%%%%%%%%%%%
%%%%%%%%%%%%%%%%%%%%%%%%%%%%%%%%%%%%%%%%%%%%%%%%%%%%%%%%%%%%%

\centerline{\null}
\vskip+3.3truein
\includegraphics{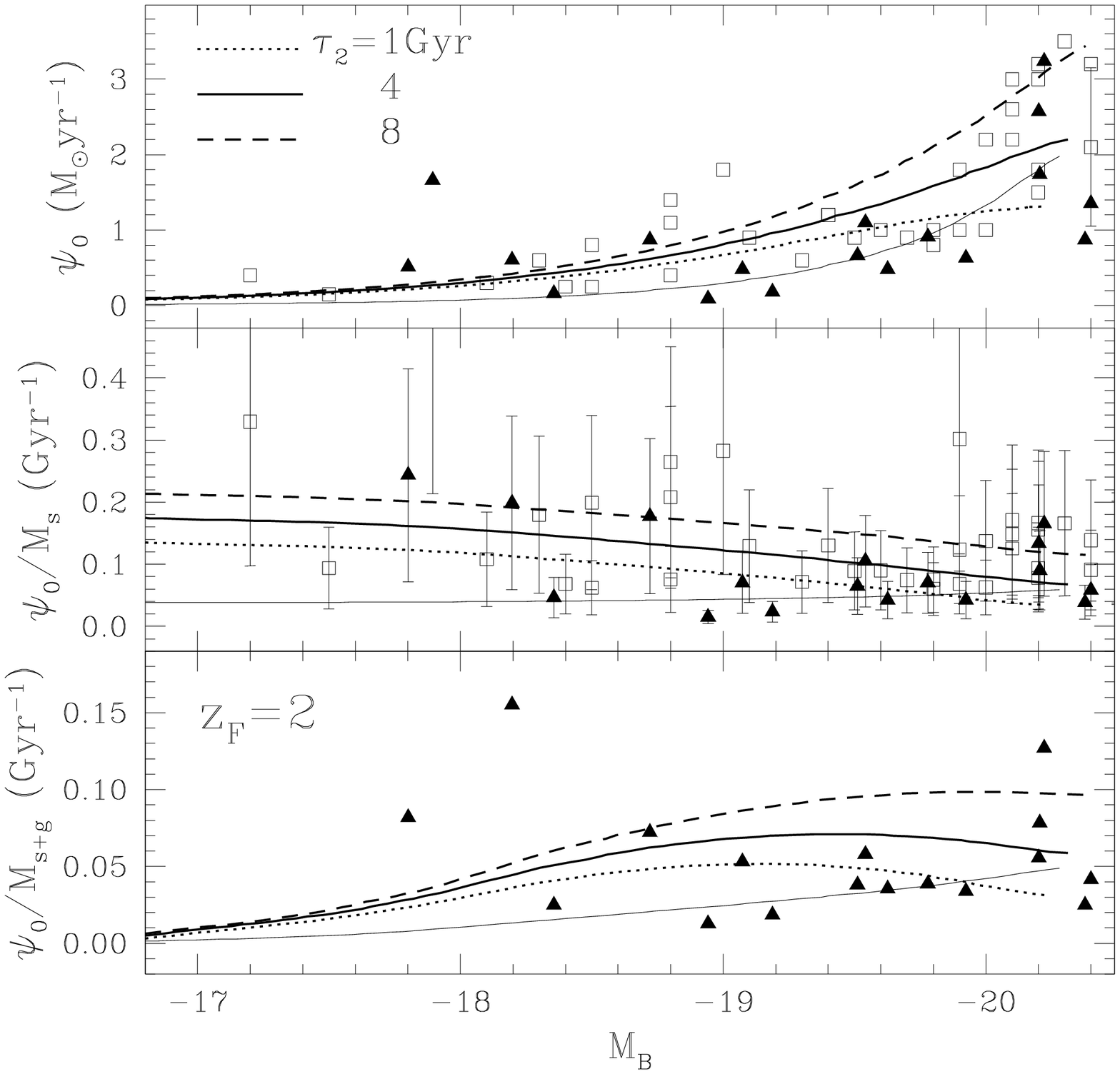}
\figcaption[fs_f7.eps]{Star formation rates at zero redshift: The
	model predictions are compared against the samples
	of Kennicutt (1983; squares) and Kennicutt (1994;
	triangles). The thick dotted, solid and dashed
	lines correspond to a linear Schmidt-type law
	with different infall timescales ($\tau_2=1, 4$ and
	$8$~Gyr, respectively). The thin solid line is the
	result for a quadratic Schmidt-type law. The
	top, medium and bottom panels compare the absolute;
	specific-weighted-by-stars, and 
	specific-weighted-by-total-mass star formation rates,
	respectively. The conversion from $B$-band absolute 
	luminosity (from the data) to stellar mass
	is done by assuming a range in mass-to-light ratios
	between $-0.15 <\log (M/L_B) < +0.15$, predicted
	by population synthesis models for a large
	range of star formation histories.	
	The points in the bottom panel 
	includes an estimate of the mass in molecular hydrogen
	using CO as tracer.\label{f7}}
\vskip+0.2truein

%%%%%%%%%%%%%%%%%%%%%%%%%%%%%%%%%%%%%%%%%%%%%%%%%%%%%%%%%%%%%

\noindent
process stars in the faintest
disk galaxies. Furthermore, the gas mass fraction:
$f_g\equiv M_g/(M_g+M_s)$, where $M_g$ and $M_s$ are the
mass in gas and stars, respectively, 
will also vary over a very large range. 
As a simple example, we can write the gas mass fraction
in a closed-box model (using $f(t)=\delta(t)$ in equation~4) as:
\begin{equation}
f_g(t) = \exp(-t/\tau_g), 
\end{equation}
with the star-processing timescale ($\tau_g$) defined in (5).
Massive systems have efficiencies close to $\ceff\sim 1$, 
i.e. $\tau_g\sim 1$~Gyr, which means the gas mass fraction
should be very low in disk galaxies at $z=0$. However,
low-mass galaxies are predicted to have $\tau_g\sim 100$~Gyr,
which boosts the gas mass fraction in local galaxies up
to 90\%. Hence, galaxies with low rotation velocities 
should have large amounts of gas. This issue is hard to 
circumvent by
redefining the model: the index of the Schmidt power 
law used in the correlation between gas mass and
star formation rates does not change the range of
efficiencies. We also considered several other models 
with a time-varying star formation efficiency but the 
final results were very similar to those presented here.

Figure~7 checks the consistency of our model with 
estimates of star formation in nearby spiral galaxies
using the flux of the H$\alpha$ emission line as
the observable. The data points are from Kennicutt
(1983, squares; 1994, triangles).
The top panel plots the prediction of the
absolute star formation rate (in $M_\odot {\rm yr}^{-1}$)
as a function of $B$-band absolute luminosity 
for three different infall timescales: 
$\tau_2=\{1, 4,$ and $8\}$~Gyr, using a linear Schmidt law. 
The thin solid line is the prediction for $\tau_2=4$~Gyr for a

%%%%%%%%%%%%%%%%%%%%%%%%%%%%%%%%%%%%%%%%%%%%%%%%%%%%%%%%%%%%%
%%%%%%%%%%%%%%%%%%%%%%%%   FIG. 8   %%%%%%%%%%%%%%%%%%%%%%%%%
%%%%%%%%%%%%%%%%%%%%%%%%%%%%%%%%%%%%%%%%%%%%%%%%%%%%%%%%%%%%%

\centerline{\null}
\vskip+3.3truein
\includegraphics{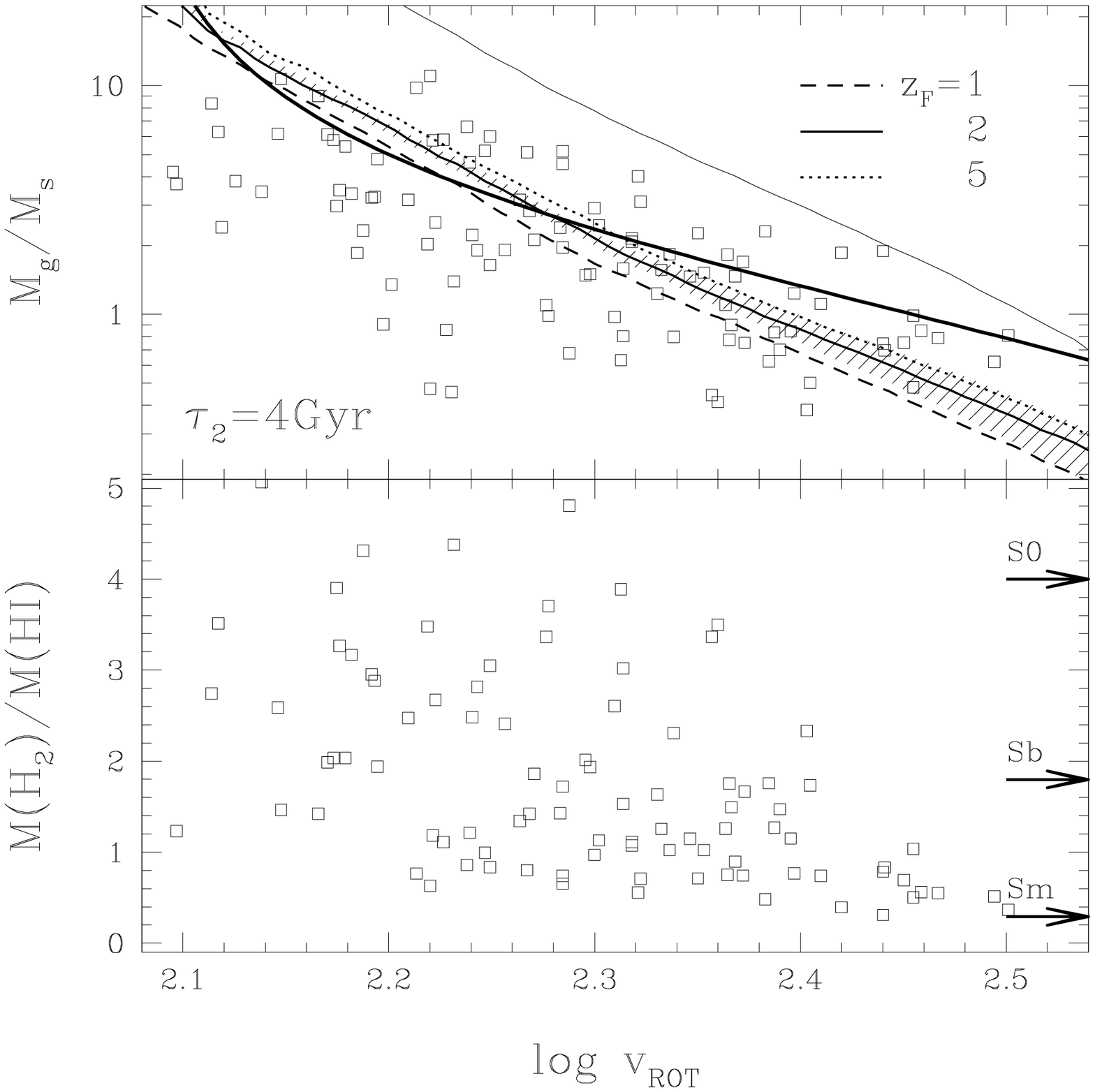}
\figcaption[fs_f8.eps]{{\sl Top panel:} Gas-to-stellar mass 
	fractions predicted by the model at three different
	formation redshifts (linear Schmidt law for $\tau_2=4$~Gyr 
	as dashed, solid and dotted lines for $z_F=1,2,$ and $5$, 
	respectively; thin solid line for $\tau_2=4$~Gyr, $z_F=2$
	for a quadratic Schmidt law). The shaded region gives the
	range of gas fractions at $z_F=2$ for $1<\tau_2/{\rm Gyr}<8$.
	The points are from McGaugh et al. (2000),
	after being corrected so that the stellar masses correspond
	to $K$-band luminosities that obey the Tully-Fisher
	relation. The thick line is the fit-by-eye of
	Bell \& de~Jong (2000) for the mass fractions found in 
	a sample of low-inclination spiral galaxies (see
	text for details). {\sl Bottom panel:} If we take the model
	predictions at face value, one has to consider
	the presence of molecular hydrogen. The points give
	the expected ratio of molecular to atomic hydrogen
	for the sample shown above. A rough upper
	limit to the gas fraction --- given by the rotation
	curves for a maximal disk --- is $\sim 6$ (Combes 2000).
	The arrows are the results from Young \& Knezek (1989)
	as a function of galaxy morphology.\label{f8}}
\vskip+0.2truein

%%%%%%%%%%%%%%%%%%%%%%%%%%%%%%%%%%%%%%%%%%%%%%%%%%%%%%%%%%%%%

\noindent
quadratic law. A typical error bar from the data
is also shown. The star formation rates obtained from 
these measurements can have uncertainties mainly due to
variable extinction up to $\pm 50$\%. The model predictions
show that even though there is not much gas in 
larger mass systems, their high star formation efficiencies
account for an increase in the SFR with luminosity.

Table~2 gives the fits to the star formation rate as 
a function of stellar mass ($\psi\propto M_s^b$) 
predicted by the model,
for various infall parameters. The power law index
lies in the range $b\sim 0.7-0.8$ and as expected, 
it varies more with formation redshift for shorter infall 
timescales. The middle panel of Figure~7 shows the
specific star formation rate weighted by stellar
mass ($\psi/M_s$), which decreases with increasing mass
since $b<1$. In order to transform the absolute luminosities
from Kennicutt (1983; 1994) to stellar masses, we 
assume a large range in mass-to-light ratios 
$-0.15<\log(M/L_B)<+0.15$ in order to account for 
different age distributions or initial mass functions 
of the stellar component.
The error bars shown give the range of specific star formation 
rates depending on the mass-to-light ratio
chosen. Finally, the bottom panel plots the specific
star formation rate weighted by total (baryonic)
matter, i.e. stars and gas. In this case only
Kennicutt (1994) gives gas masses. Both atomic and
molecular hydrogen are considered and helium
is included by correcting the final mass by a 
factor of $1.4$. The data is consistent with the
model predictions, except for the two faintest data
points, which appear to be in disagreement with such a high 
gas content. One explanation for this 
is that the molecular hydrogen measured has been
underestimated. Molecular hydrogen is usually
measured indirectly, by assuming a constant H$_2$/CO
ratio. Systems with lower metallicities than the
galaxies used to calibrate this ratio should have 
larger H$_2$ masses than the estimates according to
this technique (Combes 1999). The bottom panel
of Figure~8 also shows the results from 
Young \& Knezek (1989) regarding the dependence of
$M(H_2)/M(HI)$ with morphology, where the detection
of H$_2$ is based on CO observations. The observed trend,
which finds more molecular hydrogen in 
earlier types (which have higher metallicities),
is a reflection of the strong dependence of 
the H$_2$/CO ratio with metal abundance.

In order to explore this point further, we used the
data presented by McGaugh et al (2000) who motivated
the existence of a baryon-dominated Tully-Fisher relation,
i.e. a tight correlation between total stellar$+$gas mass
and rotational velocity. They inferred the stellar masses 
from photometric data in optical ($I$) and near infrared 
($H$ and $K^\prime$) bands, by assuming a fixed 
mass-to-light ratio (however, see Bell \& de~Jong 2001). 
Near infrared observations are usually 
good stellar mass observables since $M/L_K$ does not change 
much over a large range of ages. For consistency, we 
decided to use the Tully-Fisher relation as a constraint 
so that the stellar masses were corrected to abide 
by the linear fits shown in Table~1. The gas-to-stellar
ratios are shown in Figure~8 along with model predictions
for three formation redshifts for a linear Schmidt law
and also for a prediction using a quadratic law.
The gas fractions predicted for the quadratic Schmidt law
are higher since such a model requires even lower 
star formation efficiencies at the faint end (Figure~4).

The figure also shows the
fit-by-eye from Bell \& de~Jong (2000) who estimated
the gas fraction to be: $f_g=M_g/(M_g+M_s)=0.8 + 0.14(M_K-20)$
in a sample of low-inclination spiral galaxies. The fractions
were computed from $K$-band luminosities (in order to 
find stellar masses) and \ion{H}{1} fluxes corrected to 
include helium. They also included the contribution 
from molecular hydrogen by using the ratio of molecular to
atomic gas masses as a function of morphological type
(Young \& Knezek 1989). Our models are thereby consistent with the
data for a linear Schmidt law. However, the large scatter
observed could be associated with the presence of undetected
molecular hydrogen. The bottom panel shows the ratio
between molecular and atomic hydrogen predicted if we
take the models at face value. It is interesting
to notice that a larger amount of undetected H$_2$ is
expected for fainter galaxies, in agreement
with the fact that fainter galaxies have low efficiencies
which translate into lower average metallicities (cf fig.~2)
so that the H$_2$/CO ratio used to trace H$_2$ could be in
significant disagreement with the calibrations (Combes 2000).

The ratios shown in the figure are still consistent with
the total amount of matter to be expected from the
rotation curves of spirals. Combes (1999) showed that using
a ratio of $M(H_2)/M(HI)\sim 6.2$, the gas and stellar 
content in NGC~1560 can account for the rotation curve out 
to $8$ kpc. Furthermore, observations by Valentijn \& 
Van~der~Werf (1999) of the lowest pure rotational lines of 
H$_2$ in NGC~891 out to $11$~kpc, using the Short-Wavelength 
Spectrometer on board the Infrared Space Observatory,
found large amounts of H$_2$ at temperatures $T\sim 80-90$~K,
outweighing \ion{H}{1} by a factor $5-15$. More data is clearly 
needed to draw a definitive conclusion about the 
presence of a high mass in molecular hydrogen. Nevertheless, the 
agreement between these observations and our model predic-

%%%%%%%%%%%%%%%%%%%%%%%%%%%%%%%%%%%%%%%%%%%%%%%%%%%%%%%%%%%%%
%%%%%%%%%%%%%%%%%%%%%%%%   FIG. 9   %%%%%%%%%%%%%%%%%%%%%%%%%
%%%%%%%%%%%%%%%%%%%%%%%%%%%%%%%%%%%%%%%%%%%%%%%%%%%%%%%%%%%%%

\centerline{\null}
\vskip+3.3truein
\includegraphics{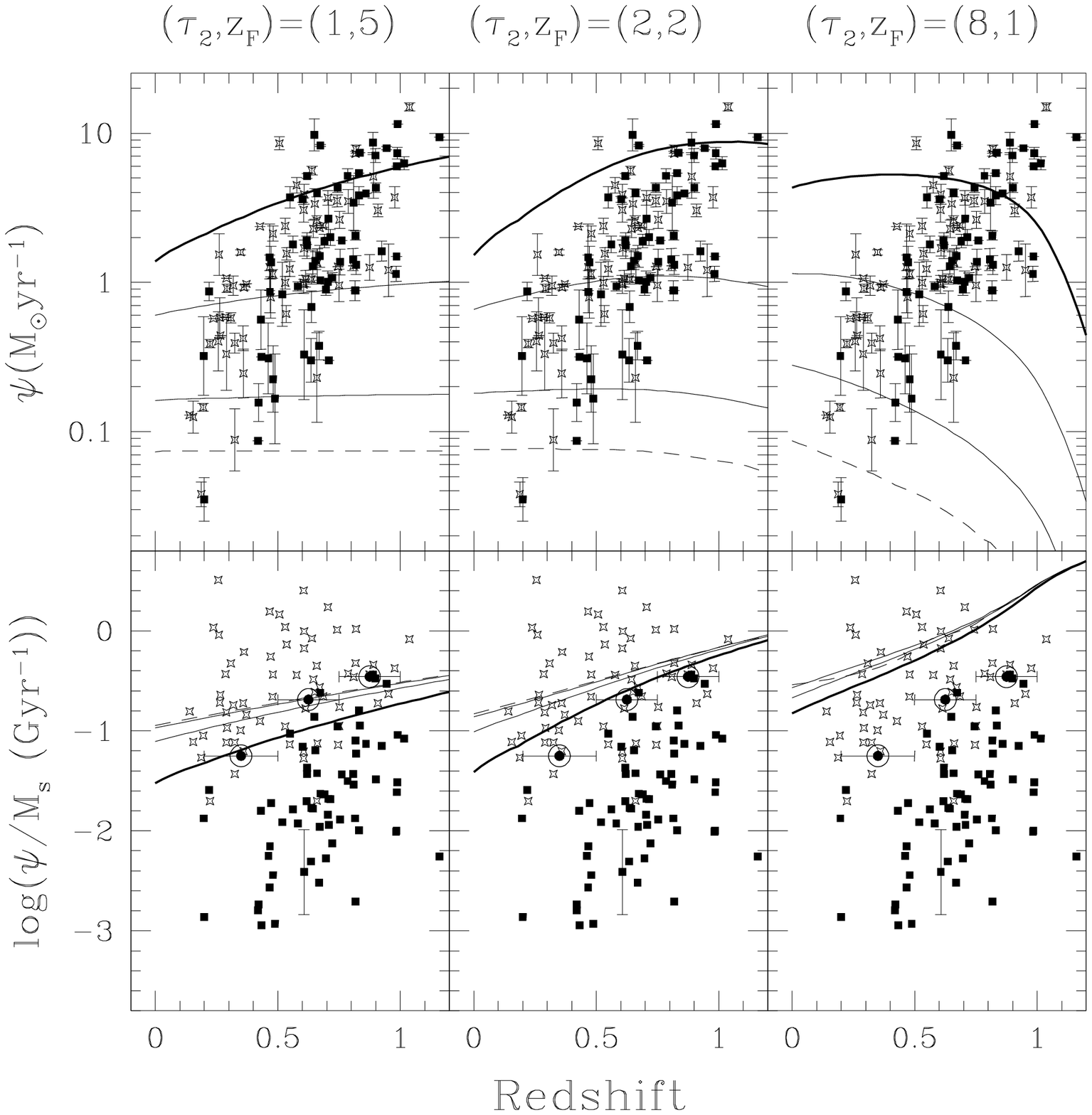}
\figcaption[fs_f9.eps]{Redshift evolution of the absolute ({\sl top}) 
	and specific star formation rate weighted by stellar 
	mass ({\sl bottom}), for a set of rotational velocities:
	The curves correspond to $\log v_{\rm ROT}=2.6,$ (thick) 
	$ 2.4, 2.2$ and $2.0$ (dashed line). The infall parameters
	$(\tau_2/{\rm Gyr},z_F)$ are shown on top. The data 
	points are from the sample of Brinchmann (1999). The error bars
	come straight from observational uncertainties of 
	[\ion{O}{2}] emission, used as tracer of star formation. 
	The sample has been divided with respect to stellar mass
	at the median ($\log (M_s/M_\odot)=10.3$), with high-mass
	galaxies represented as filled squares. In the bottom 
	panels we only show one typical error bar for clarity 
	purposes. The circles with horizontal error bars
	give the best fit for the specific star formation rate
	at $\log(M_s/M_\odot)=10$ found by Brinchmann \& Ellis 
	(2000) in three redshift bins over the redshift range 
	represented by the error bars.\label{f9}}
\vskip+0.2truein

%%%%%%%%%%%%%%%%%%%%%%%%%%%%%%%%%%%%%%%%%%%%%%%%%%%%%%%%%%%%%

\noindent
tions --- arising from the need for very low star formation efficiencies
in low-mass galaxies --- motivates further exploration of this
problem, especially given the impact that such an
issue has on galaxy formation as well as on our understanding
of the properties of dark matter.

We want to emphasize that a possible weakness in the model 
presented here is the {\sl absolute} estimate of star formation 
efficiencies. One should take the values of $\ceff$ as lower
bounds for a given galaxy if the results are to be compared
with local efficiencies defined as 
$A\sim\Sigma_\psi/\Sigma^n_g$, where $\Sigma_g$ and $\Sigma_\psi$ 
are the surface densities of gas and star formation rate, 
respectively. However, even this method of estimating efficiencies
is questionable after the discovery of star formation in the
extreme outer regions of disk galaxies, which cannot be
described by a single-component Schmidt law (Ferguson et al. 1998).
The stellar metallicities predicted by our model for rotation 
velocities corresponding to a Milky Way sized galaxy (which 
implies $0.1\simlt\ceff\simlt 0.2$) are in the right
ballpark: $-0.1<[{\rm Fe/H}]<+0.05$, whereas low efficiencies
$\ceff\sim 0.01-0.02$ predicted by the model for a galaxy
like the Large Magellanic Cloud ($M_B=-18$) give metallicities
($-1.1<[{\rm Fe/H}]<-0.8$) that are consistent with the observations 
of the young clusters --- with ages $t\simlt 3$~Gyr --- with 
the lowest metallicities (Olszewski, Suntzeff \& Mateo 1996).
However, notice the large range of metallicities found in old
LMC clusters ($-2.1<[{\rm Fe/H}]<-1.4$ for $t\sim 14$~Gyr) which 
hints at a more complicated star formation history.

%%%%%%%%%%%%%%%%%%%%%%%%%%%%%%%%%%%%%%%%%%%%%%%%%%%%
%%%      Table 2. Fits to SFRs                   %%%
%%%%%%%%%%%%%%%%%%%%%%%%%%%%%%%%%%%%%%%%%%%%%%%%%%%%

 \begin{table*}
   \begin{center}
   \caption{Star Formation Rates:
	$\log(\psi/M_\odot yr^{-1})={\cal A} + b\log(M_s/M_\odot)$}
   \label{table-2}
   \vskip+0.1truein
   \begin{tabular}{cc|cc}\hline\hline
  $\tau_2$/Gyr & $z_F$ & $b$ & ${\cal A}$\\  
  \hline
  $1$ & $1$ & $0.665$ & $-6.763$ \\
      & $2$ & $0.716$ & $-7.311$ \\
      & $5$ & $0.741$ & $-7.597$ \\
  $4$ & $1$ & $0.796$ & $-7.817$ \\
      & $2$ & $0.795$ & $-7.937$ \\
      & $5$ & $0.796$ & $-8.018$ \\
  $8$ & $1$ & $0.859$ & $-8.311$ \\
      & $2$ & $0.862$ & $-8.449$ \\
      & $5$ & $0.862$ & $-8.532$ \\
  \hline\hline
   \end{tabular}
   \end{center}
 \end{table*}

%%%%%%%%%%%%%%%%%%%%%%%%%%%%%%%%%%%%%%%%%%%%%%%%%%%%%%%%%%%%%%%%%%%%%
\section{Evolution of the Tully-Fisher relation with redshift}

The star formation histories given by the best fits to the 
efficiency and the fraction of gas in outflows shown in 
Figures~4 and 5 can be used to evolve the systems backwards 
in time. A major consequence of the large range in star 
formation efficiencies found with respect to rotational 
velocity is that low-mass systems will have a broader 
distribution of stellar ages. This implies a stronger 
evolution in luminosity with lookback time. Furthermore, a low 
efficiency will yield a constant and low star formation rate 
with redshift. On the other hand, massive galaxies will undergo 
milder changes in absolute luminosity because of their high star 
formation efficiencies, and this will narrow the expected 
age distribution (cf. Fig.~3). 
The star formation rate for massive galaxies, 
however, experiences a significant increase with redshift. 
Figure~9 illustrates this point: model predictions of the 
redshift evolution of the absolute ({\sl top}) and specific 
weighted-by-mass ({\sl bottom}) star formation 
rates are shown for $\log v_{\rm ROT} = \{ 2.6$ (thick line),
$2.4, 2.2,$ and $2.0$ (dashed)$\}$. An increase in the specific 
star formation rate with decreasing mass is predicted, 
spanning a factor of $4$ in $\psi/M_s$ between 
$\log v_{\rm ROT}=2.0$ and $2.6$, i.e. roughly linear
with $v_{\rm ROT}$. This difference in $\psi/M_s$ gets
narrower at high redshifts, explained by the fact that 
the increase in the specific SFR in galaxies with a high 
efficiency will evolve faster with lookback time, whereas
low-mass systems --- with low values of $\ceff$ --- feature a
low and nearly constant SFR.

The estimates of stellar mass and star formation rates of a sample 
of field galaxies from the Canada France Redshift Survey and
from the Hubble Deep Field North and its flanking fields are
also shown in Figure~8 (Brinchmann 1999). The sample spans a
wide redshift range ($0.2<z<1$). The stellar masses 
are obtained following $K$-band luminosities, but adding a 
correction term based on a comparison between multiband 
optical photometry and population synthesis models for various
ages, metallicities and dust extinction values (Brinchmann 
\& Ellis 2000). These authors estimated the uncertainties in the
computation of the stellar mass to be $\log\Delta M_s\sim 0.3$~dex.
The ongoing star formation rate is computed 
from the flux of the [\ion{O}{2}] emission line. The sample is divided 
in Figure~8 with respect to stellar mass, so that galaxies 
with masses larger (smaller) than the median 
($\log(M_s/M_\odot)=10.3$) are represented by squares (stars). 
The bottom panels only show one typical error bar to avoid 
crowding the figure. 
Taking the model predictions at face value,
one would explain the lack of data points in the bottom right
portion of the top panels by a selection bias of the sample 
which would have to be missing systems with low rotational velocities 
(or faint absolute luminosities) at high redshift. It  is
also interesting to notice that the region with 
$\psi\simgt 5 M_\odot {\rm yr}^{-1}$  is mainly populated by
massive galaxies. Brinchmann \& Ellis (2000) find a significant
increase in the specific star formation rate with redshift, shown
in the bottom panels of Figure~9 as big circles with error bars.
The points are the values at $\log (M_s/M_\odot )=10$ of the 
best fit found in the redshift bins represented by the
horizontal error bars. The values for higher (lower) mass
systems will roughly shift linearly to lower (higher) specific
SFRs. Hence, this fits into our model by assuming a correlation
between stellar mass and infall parameters, so that galaxies
with more extended star formation histories should correspond
to low-mass systems and vice versa.

The redshift evolution of the Tully-Fisher relation is another
interesting issue which becomes possible to quantify with recent
observations of the rotation curves of disk galaxies at moderate
and high redshift (Vogt et al. 1997). The power of an analysis
including star formation lies in the ability to predict relative
changes in luminosity evolution caused by a different age 
distribution of the stellar populations. In contrast to semi-analytic
modellers who are more concerned with the evolution of the zero point of
the Tully-Fisher relation, we can ``turn the crank'' of our 
model --- which is constrained at zero redshift by the local
Tully-Fisher relation --- backwards in time and find luminosity 
changes as a function of galaxy mass. Figure~10 shows that
even more importantly than the zero point, {\sl the slope} 
($\gamma$, with $L\propto v_{\rm ROT}^\gamma$) of the
Tully-Fisher relation is a crucial parameter to explore. 
The panels show the evolution of several properties in 
optical ($B$; left) and near infrared ($K$; right) passbands.
The bottom panels show the predicted evolution
of $\gamma_{B,K}$ for $\tau_2=4$~Gyr and three formation redshifts 
using a linear Schmidt law, and for $z_F=2$ with a 
quadratic law (dashed-dotted line). The shaded areas in all panels
give the predictions at $z_F=2$ when allowing a wide range
of infall timescales ($1 < \tau_2/{\rm Gyr} < 8$).
The steepening of the slope
is a consequence of the very low star formation efficiencies
at the low-mass end, which imply a broader distribution
of stellar ages. 
Models based on simple arguments of structure formation fail
to include the effect of the evolution of the stellar populations.
Hence, the prediction of Mo, Mao \& White (1998), who tie the
evolution of the luminous component in disks to the
parameters defining the dark matter halo in which they are embedded,
uses a scaling argument to find a constant $\gamma =3$,  
obviously regardless of passband and therefore inconsistent with
the observations. 

A slope change implies that the measured zero point will depend on
the absolute luminosity considered. The top panels plot the
luminosity evolution for systems with a rotational velocity
of $100$ km s$^{-1}$, whereas the middle panels correspond to
massive galaxies, with $\log v_{\rm ROT}=2.6$. As expected, 
more massive disks undergo a milder luminosity evolution. 
Recent observations at moderate and high redshift are shown 
as well. The latest work of Vogt et al. (2000; filled squares) 
on disks in the Groth Survey Strip, extending over a large 
redshift range: $0.2<z<1$, shows a very small
--- if any --- evolution in absolute $B$-band luminosity
($\Delta M_B\simlt -0.2$). The observations of Bershady et al. 
(1999; hollow squares) at $z\sim 0.4$ hint at a slightly different luminosity 
evolution depending on rest frame $B-R$ color, which we coarsely
map into $\log v_{\rm ROT}$. Hence, low-mass systems
are found to have $\Delta M_B=-0.5\pm 0.5$ whereas massive disks
appear not to have any significant evolution: $\Delta M_B=0.0\pm 0.5$.
This result is in good agreement with the model,
which gives a larger luminosity evolution for low-mass
disks. The triangle represents the result from 
Simard \& Pritchet (1998), who find very strong evolution
($\Delta M_B\sim -2$) in a sample of strong [\ion{O}{2}] emitters 
at $z\sim 0.35$, in remarkable contrast with the other observations
as well as with our model. One could speculate 
that disks with a star formation rate peaked at $z_F\sim 0.5-1.0$
could in principle explain this last data point, assuming that the
selection criterion biased the sample towards systems with
a very recent star formation history.

%%%%%%%%%%%%%%%%%%%%%%%%%%%%%%%%%%%%%%%%%%%%%%%%%%%%%%%%%%%%%%%%%%%%%
\section{Discussion}

In this paper we have presented a phenomenological approach to galaxy
formation which reduces the process of star formation to a set
of a few parameters controlling the infall of primordial gas,
the efficiency of processing gas to form stars, and the fraction
of gas and metals that are ejected from the galaxy. These
parameters are not set {\sl a priori}. Instead, they are allowed 
to vary over a large volume of parameter space. The Tully-Fisher (1977)
relation in different wavebands is used as a constraint, in order 
to find the parameter space compatible with the data. Our results 
show that gas outflows should not play a major role in explaining 
the color and luminosity range of disk galaxies, whereas 
a very large range of star formation efficiencies ($\ceff$)
is required. A power law fit to the correlation between $\ceff$
and rotational velocity ($v_{\rm ROT}$) gives a slope 
$\Delta\log\ceff /\Delta\log v_{\rm ROT}\sim 4$ regardless
of either the infall parameters chosen or the exponent used
in the Schmidt law relating gas density and star formation
rates (see Figure~4). For a range of rotational velocities
$2.1<\log v_{\rm ROT}<2.6$ --- which maps into a $4.5$ mag
range in absolute $K$-band luminosity --- we therefore
predict a star formation efficiency variation of around $2$
orders of magnitude. This corresponds to star processing
timescales which are $100$ times longer for low-mass disks
compared to the bright ones. In contrast, Bell \& Bower 
(2000) suggest a mass dependency of either infall or
outflows in order to explain the observations, although 
their models did not allow for a wide range of star
formation efficiencies. Moreover, Boissier et al. (2001) 
found no correlation between the star formation efficiency
and the galaxy mass, using a chemo-spectrophotometric
evolution model calibrated on the Milky Way and extended
to a wide range of galaxy masses. However, the observed color 
range must be accounted for by a large range of formation redshifts, so
that massive disks are formed {\sl earlier} than less massive ones.
The predicted large variation in stellar ages as a function

%%%%%%%%%%%%%%%%%%%%%%%%%%%%%%%%%%%%%%%%%%%%%%%%%%%%%%%%%%%%%
%%%%%%%%%%%%%%%%%%%%%%%%   FIG. 10  %%%%%%%%%%%%%%%%%%%%%%%%%
%%%%%%%%%%%%%%%%%%%%%%%%%%%%%%%%%%%%%%%%%%%%%%%%%%%%%%%%%%%%%

\centerline{\null}
\vskip+3.3truein
\includegraphics{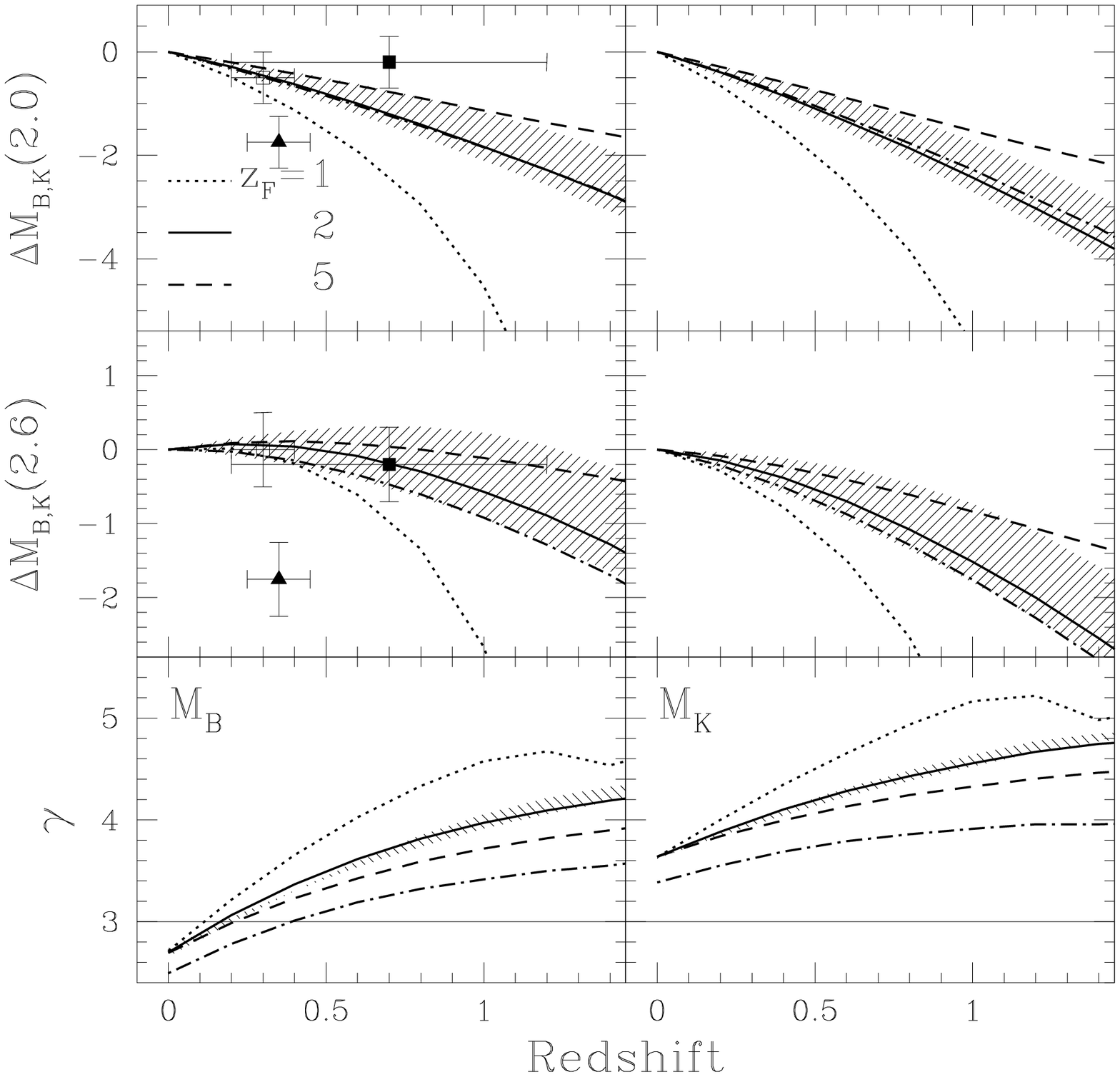}
\figcaption[fs_f10.eps]{Redshift evolution of the Tully-Fisher relation. 
	The evolution with redshift of the zero points at 
	$\log v_{ROT}=2$ ({\sl top panels}) and $2.6$ 
	({\sl medium panels}) are shown as well as the change in
	slope $\gamma$ ($L\propto v_{ROT}^\gamma$) 
	in optical ($B$; {\sl left}) and near-infrared 
	bands ($K$; {\sl right}). The thick dotted, solid and dashed
	lines correspond to formation redshifts of $z_F=1,2$
	and $5$, respectively, and all three lines assume 
	an infall timescale $\tau_2=4$~Gyr and a linear Schmidt law. 
	The predictions for a quadratic law, with $\tau_2=4$~Gyr 
	and $z_F=2$ are shown as a dash-dot line. The shaded area 
	gives the predictions for a range in infall timescales,
	between $\tau_2=1$ and $8$ Gyr at $z_F=2$. The points
	show the offset of the Tully-Fisher relation in rest 
	frame $B$ band, in observations at moderate and high 
	redshift from Bershady et al. (1999; hollow square); 
	Vogt et al. (2000; filled square), and Simard \& 
	Pritchet (1998; triangle). The horizontal error bar just
	encompasses the redshift range of the observations. In the
	bottom panels, the horizontal line at $\gamma =3$ shows the
	prediction of a simple dynamical model that does not
	consider the evolution of the stellar populations
	(e.g. Mo, Mao \& White 1998).\label{f10}}
\vskip+0.2truein

%%%%%%%%%%%%%%%%%%%%%%%%%%%%%%%%%%%%%%%%%%%%%%%%%%%%%%%%%%%%%

\noindent
of galaxy mass may be in conflict with the redshift evolution 
of both the luminosity function of disk galaxies and the 
Tully-Fisher relation. If we impose on our models formation redshifts 
$z_F\simgt 1$ for all galaxy masses, a large range of efficiencies 
are predicted. Furthermore, a hierarchical scenario --- for which
small structures collapse first --- would increase the slope
of the correlation between $\ceff$ and rotation velocity.
The departure of the scaling behavior of the star formation 
efficiency found in this paper with respect
to dynamical back-of-the-envelope estimates frequently
used in semi-analytic modelling, i.e. 
$\ceff\propto t_{dyn}^{-1}\propto v_{\rm ROT}^3/M$, 
where $t_{dyn}$ is a characteristic 
dynamical time-scale for the galaxy (eg. Kauffmann, White 
\& Guiderdoni 1993), shows that the Tully-Fisher relation
cannot be fully explained by the properties of the 
halo, but rather, needs a component incorporating self-regulated
star formation (Silk 1997, 2000).

We also applied our model using the Faber-Jackson (1976)
correlation in spheroids as a constraint, finding 
a rather constant and high efficiency and a significant
range in gas outflows. This behavior could be explained by the
different dynamical history of disks and spheroids. Early-type
systems have undergone mergers with equal mass progenitors, which
can drive a significant amount of gas out of galaxies. On the
other hand, the standard framework for disk galaxy formation
assumes smooth infall and cooling of gas on to the centers
of dark matter halos. Furthermore, the low outflow fractions 
found in disks for any galaxy mass enable us to reject 
supernova-driven winds as an important
mechanism for driving gas and metals out of the interstellar medium
of galaxies, in agreement with the hydrodynamic simulations of
McLow \& Ferrara (1999) and the modelling of star formation
in dwarf galaxies by Ferrara \& Tolstoy (2000).

The large range of star formation efficiencies obtained by our
model implies that low-mass galaxies should have a large fraction of
baryonic matter in the form of gas. As a sanity check, we compared 
the predictions for the star formation rate at $z=0$ as a 
function of stellar mass with Kennicutt (1983; 1994) and found
good agreement. A fit to a power law --- $\psi\propto M_s^b$ ---
gives an exponent $b\sim 0.7-0.8$ for a large range of infall
parameters (Table~2). Hence, in the tug-of-war between star
formation efficiency and galaxy masses, the former wins over,
giving higher star formation rates in brighter galaxies.
In a speculative mode, we suggest that a
large fraction of the considerable amounts of gas predicted for 
systems with low efficiencies could be in cold molecular 
hydrogen, a rather elusive component. In the bottom panel 
of Figure~8 we took our model at face value and compared it 
with the stellar and gas content (atomic hydrogen plus helium) 
from the sample compiled by McGaugh et al. (2000), and we inferred 
a significant increase in the ratio of molecular to atomic
hydrogen for less massive galaxies. This can be accounted for if
there is a significant offset in galaxies with low metallicities 
from the usual H$_2$ to CO conversion factor used to detect 
molecular hydrogen (Combes 2000). Using a direct technique to
search for H$_2$, large amounts
of not-too-cold molecular gas ($T\sim 80-90$~K) have been detected 
through the lowest pure rotational transitions in NGC~891
(Valentijn \& Van~der~Werf 1999). Incidentally, an extrapolation
of up to $M(H_2)/M(HI)\sim 6.2$ can account for the rotation curves
of disks (Combes 1999). This issue has important implications not 
only for galaxy formation but also for cosmology: the existence 
of such large amounts of H$_2$ on galaxy scales could favor a 
parameter space of dark matter with partially suppressed 
hierarchical galaxy formation on sub-galactic scales, as
a consequence of the enhanced free streaming lengths found 
in scenarios such as warm dark matter (e.g. Col\'\i n, 
Avila-Reese \& Valenzuela 2000; Bode, Ostriker \& Turok 2000).
We remark that the large range in star formation efficiencies should
be incorporated into models of galaxy formation. Semi-analytical
simulations of galaxy formation which assume a constant star
formation efficiency as a function of circular velocity
(Lacey \& Silk 1991; White \& Frenk 1991; Cole et al. 2000; 
Kauffmann et al. 1999; Somerville \& Primack 1999) 
are far too simplistic to give meaningful predictions of, 
for example, the cosmic star formation history.

We used our model to infer the redshift evolution 
of star formation and of the Tully-Fisher relation. 
Star formation rate predictions are compatible with
the data, although the scatter, uncertainties and various
biases makes this a rather uncertain endeavor. SFRs are
expected to undergo stronger evolution in systems with high
rotational velocities, whereas low-mass systems feature
a steadier formation rate due to their low efficiencies. 
The range of specific formation rates with stellar mass is
predicted to decrease with redshift. The Tully-Fisher
relation is expected to evolve {\sl in slope}, as well as in
zero point, because the age distribution of the stellar population
varies with galaxy mass, being more spread out in low-mass disks.
Hence, we expect a steepening of the slope. For instance, in 
rest frame $B$ band the slope changes from $\gamma_B\sim 3$ 
at zero redshift ($L\propto v_{\rm ROT}^\gamma$) up to 
$\gamma\sim 4-4.5$ at $z\sim 1$ (Figure~10), with a slight dependence 
on the formation redshift, giving stronger evolution for low 
formation redshifts.
The zero point should thus be given with respect to
some fiducial rotational velocity. We show the evolution of the
zero point at low ($\log v_{\rm ROT}=2$) and high mass ($2.6$) is
in agreement with the latest observations from Vogt et al. (2000)
for high mass systems, as well as with Bershady et al. (1998)
both for high and low masses. Notwithstanding the many 
observational difficulties that complicate accurate estimates
of the redshift evolution of $\gamma$, we find the evolution of
the slope of the Tully-Fisher relation to be the best observable
for exploring the star formation process in disk galaxies,
in analogy with the study of the slope of the fundamental plane
in early-type systems (Ferreras \& Silk 2000b).

\section*{Acknowledgments}
IF is supported by a grant from the European Community under 
contract HPMF-CT-1999-00109. We would very much like to thank
Jarle Brinchmann for helpful discussions and for making 
available the data from his PhD thesis. We would also 
like to thank Eric Bell and Stacy McGaugh for providing 
their data on disk gas masses. The anonymous referee is
gratefully acknowledged for useful comments and suggestions.


\begin{references}
\reference{ba98} Baugh, C.~M., Cole, S., Frenk, C.~S. \& 
	Lacey, C.~G. 1998, \apj, 498, 504
\reference{bb00} Bell, E.~F. \& Bower, R.~G. 2000, \mnras, 319, 235
\reference{bdj00} Bell, E.~F. \& de~Jong, R.~S. 2000, \mnras, 312, 497
\reference{bdj01} Bell, E.~F. \& de~Jong, R.~S. 2001, \apj, March 20, 
	astro-ph/0011493
\reference{ber99} Bershady, M.~A., Haynes, M.~P., Giovanelli, R. \&
	Andersen, D.~R. 1999, in Merritt, D.~R., Valluri, M. \&
        Sellwood, J.~A., eds, ASP Conf. Ser. 182, Galaxy Dynamics.
	Astron. Soc. Pac., San Francisco, p. 499
\reference{bot00} Bode, P., Ostriker, J.~P. \& Turok, N. 2000, astro-ph/0010389
\reference{boi01} Boissier, S., Boselli, A., Prantzos, N. \&
	Gavazzi, G. 2001, \mnras, 321, 733
\reference{ble92} Bower, R.~G., Lucey, J.~R. \& Ellis, R.~S., 
	1992, \mnras, 254, 589
\reference{br99} Brinchmann, J. 1999, PhD thesis, Univ. Cambridge
\reference{br00} Brinchmann, J. \& Ellis, R.~S. 2000, \apj, 536, L77
\reference{cwb96} Charlot, S., Worthey, G. \& Bressan, A. 1996, \apj, 457, 625
\reference{co00} Cole, S., Lacey, C.~G., Baugh, C.~M. \& Frenk, C.~S. 2000, 
	\mnras, 319, 168
\reference{cav00} Col\'\i n, P., Avila-Reese, V. \& Valenzuela, O.
	2000, \apj, 542, 622
\reference{co00} Combes, F. 2000, astro-ph/0007318
\reference{co99} Combes, F. 1999, astro-ph/9910296
\reference{deg99} de~Grijs, R. \& Peletier, R.~F. 1999, \mnras, 310, 157
\reference{fj76} Faber, S.~M. \& Jackson, R. 1976, \apj, 204, 668
\reference{fer98} Ferguson, A.~M.~N., Wyse, R.~.F.~G., Gallagher, J.~S.
	\& Hunter, D.~A. 1998, \apj, 506, L19
\reference{ft00} Ferrara, A. \& Tolstoy, E. 2000, \mnras, 313, 291
\reference{fs00a} Ferreras, I. \& Silk, J. 2000a, \apj, 532, 193
\reference{fs00b} Ferreras, I. \& Silk, J. 2000b, \mnras, 316, 786
\reference{gio97} Giovanelli, R., Haynes, M.~P., Herter, T., 
	Vogt, N.~P., Wegner, G., Salzer, J.~J., 
	Da~Costa, L.~N. \& Freudling, W. 1997, \aj, 113, 22
\reference{gn98} Gnedin, N.~Y. 1998, \mnras, 294, 407
\reference{iba95} Ibata, R.~A. \& Gilmore, G.~F. 1995, \mnras, 275, 605
\reference{jc92} Jacoby, G.~H. et al. 1992, \pasp, 104, 599
\reference{kwg93} Kauffmann, G., White, S.~D.~M. \& Guiderdoni, G.
	1993, \mnras, 264, 201
\reference{ka99} Kauffmann, G., Colberg, J.~M., Diaferio, A. \&
	White, S.~D.~M. 1999, \mnras, 303, 188
\reference{ken83} Kennicutt, R.~C. 1983, \apj, 272, 54
\reference{ken94} Kennicutt, R.~C., Tamblyn, P. \& Congdon, C.~W.
	1994, \apj, 435, 22
\reference{ken98a} Kennicutt, R.~C. 1998a, \araa, 36, 189
\reference{ken98b} Kennicutt, R.~C. 1998b, \apj, 498, 541
\reference{kod98} Kodama, T., Arimoto, N., Barger, A.~J. \&
	Arag\'on-Salamanca, A. 1998, \aap, 334, 99
\reference{ls} Lacey, C. \& Silk, J. 1991, \apj, 381, 14
\reference{lar74} Larson, R.~B. 1974, \mnras, 169, 229
\reference{li95} Lilly, S.~J., Tresse, L., Hammer, F., Crampton, D.
	\& Le~F\`evre, O. 1995, \apj, 455, 108
\reference{mcg00} McGaugh, S.~S., Schombert, J.~M., Bothun, G.~D. 
	\& de Blok, W.~J.~G. 2000, \apj, 533, L99
\reference{mf99} MacLow, M.-M. \& Ferrara, A. 1999, \apj, 513, 142
\reference{mf96} Mathewson, D.~S. \& Ford, V.~L. 1996, \apjs, 107, 97
\reference{mmw98} Mo, H.~J., Mao, S. \& White, S.~D.~M. 1998, \mnras, 295, 319
\reference{osm96} Olszewski, E.~W., Suntzeff, N.~B. \& Mateo, M. 
	1996, \araa, 34, 511
\reference{ri90} Rich, R.~M. 1990, \apj, 362, 604
\reference{rpm96} Rocha-Pinto, H.~J. \& Maciel, W.~J. 1996, 
	\mnras, 279, 447
\reference{sal55} Salpeter, E.~E. 1955, \apj, 121, 161
\reference{sca86} Scalo, J.~M. 1986, fund. Cosm. Phys., 11, 1
\reference{scha92} Schaller, G., Schaerer, D., Maeder, A. \& Meynet, G. 
	1992, \aaps, 96, 269
\reference{sch59} Schmidt, M. 1959, \apj, 129, 243
\reference{si97} Silk, J. 1997, \apj, 481, 703
\reference{si00} Silk, J. 2000, \mnras, in press, astro-ph/0010624
\reference{sp98} Simard, L. \& Pritchet, C.~J. 1998, \apj, 505, 96
\reference{sp99} Somerville, R.~S. \& Primack, J.~R. 1999, \mnras, 310, 1087
\reference{ti80} Tinsley, B.~M. 1980, Fund. Cosm. Phys., 5, 287
\reference{tf77} Tully, R.~B. \& Fisher, J.~R. 1977, \aap, 54, 661
\reference{tul96} Tully, R.~B., Verheijen, M.~A.~W., Pierce, M.~J., 
	Huang, J.-S. \& Wainscoat, R.~J. 1996, \aj, 112, 2471
\reference{val99} Valentijn, E.~A. \& van der Werf, P. 
	1999, \apj, 522, L29
\reference{ver97} Verheijen, M.~A.~W. 1997, PhD thesis, Univ. Groningen
\reference{vogt97} Vogt, N.~P., et al. 1997, \apj, 479, L121
\reference{vogt00} Vogt, N.~P. 2000, in Combes, F., Mamon, G.~A. \&
  	Charmandaris, V., eds, ASP Conf. Ser. 197, Dynamics of
	Galaxies: from the Early Universe to the Present. 
	Astron. Soc. Pac., San Francisco, p. 435
\reference{wr78} White, S.~D.~M. \& Rees, M.~J. 1978, \mnras, 183, 341
\reference{wf91} White, S.~D.~M. \& Frenk C.~S. 1991, \apj, 379, 52
\reference{wdj96} Worthey, G., Dorman, B. \& Jones, L.~A., 1996, 
	\aj, 112, 948
\reference{yk89} Young, J.~S. \& Knezek, P.~M. 1989, \apj, 347, L55
\reference{yo96} Young, J.~S., Allen, L., Kenney, J.~D.~P., Lesser, A.
	\& Rownd, B. 1996, \aj, 112, 1903
\reference{zar94} Zaritsky, D., Kennicutt, R.~C. \& Huchra, J.~P.
	1994, 420, 87
\end{references}
\end{document}